\documentclass[twocolumn]{aastex63}

\usepackage{graphicx}   
\usepackage{subfigure}
\usepackage{multirow}
\usepackage{amsmath}
\usepackage{lineno}

\graphicspath{{./}{figures/}}
\shorttitle{width of cosmic filaments}
\shortauthors{Zhu, Wang, Zhang, Zheng \& Feng.}

\begin{document}
\title{Relation between the local width and linear halo mass density of cosmic filaments}

\correspondingauthor{Weishan Zhu}
\email{zhuwshan5@mail.sysu.edu.cn}

\author[0000-0002-1189-2855]{Weishan, Zhu}
\affil{School of Physics and Astronomy, Sun Yat-Sen University, Zhuhai campus, No. 2, Daxue Road \\
Zhuhai, Guangdong, 519082, China}

\author{Tian-Rui, Wang}
\affil{School of Physics and Astronomy, Sun Yat-Sen University, Zhuhai campus, No. 2, Daxue Road \\
Zhuhai, Guangdong, 519082, China}

\author{Fupeng, Zhang}
\affil{School of Physics and Materials Science, Guangzhou University\\
Guangzhou, Guangdong, 510006, China}

\author{Yi, Zheng}
\affil{School of Physics and Astronomy, Sun Yat-Sen University, Zhuhai campus, No. 2, Daxue Road \\
Zhuhai, Guangdong, 519082, China}

\author{Long-Long, Feng}
\affil{School of Physics and Astronomy, Sun Yat-Sen University, Zhuhai campus, No. 2, Daxue Road \\
Zhuhai, Guangdong, 519082, China}



\begin{abstract}
Large-scale cosmic filaments may have played an important role in shaping the properties of galaxies. Meanwhile, cosmic filaments are believed to harbor a substantial portion of the missing baryons at redshift $z<2$. To inspect the role of filaments in these issues, many properties of filaments need to be examined, including their lengths, thicknesses, and density profiles. However, measuring some of these properties poses challenges. This study concentrates on estimating filament width/thickness, investigating potential correlations between the local width of filaments and the properties of dark matter halos within filaments. We find that the local width of filaments generally increases with the mass of dark matter halos embedded in filaments per unit length, roughly following a second-order polynomial, although with notable scatter. We probe and discuss means that may refine our findings. After further verification and improvements, this relation could be applied to filament samples constructed from the observed galaxy distribution, aiding in understanding the roles of cosmic filaments in galaxy evolution and uncovering the missing baryons. 

\end{abstract}



\section{Introduction} \label{sec:intro}

Over the last 50 years, galaxy surveys have revealed a web-like spatial distribution of galaxies(e.g. \citealt{1986ApJ...302L...1D,2003astro.ph..6581C,2014MNRAS.438..177A,2014MNRAS.438.3465T}), consisting of voids, sheets, filaments and nodes. This discovery validates predictions from theoretical studies regarding the anisotropic gravitational collapse of cosmic matter predicted the (e.g. \citealt{1970A&A.....5...84Z,1996Natur.380..603B,2008LNP...740..335V}). In recent years, the evolution of the cosmic web has also been explored in great detail, facilitated by the development of numerous tools tailored for analyzing the cosmic web constructed from observed galaxies and simulations (e.g. \citealt{2005MNRAS.359..272C,2007A&A...474..315A,2009MNRAS.396.1815F,2010ApJ...723..364A,2010MNRAS.408.2163A,2012MNRAS.425.2049H,2014MNRAS.441.2923C,2018MNRAS.473.1195L,2023arXiv231101443H}). Meanwhile, several important problems arise, including the properties of the four types of structures within the cosmic web across various underground cosmology models (e.g. \citealt{2006MNRAS.366.1201N,2010MNRAS.403.1392L,2011IJMPS...1...41V,2018MNRAS.479..973C}), the impact of the cosmic web environment on the properties of dark matter halos and galaxies (e.g., \citealt{2007MNRAS.375..489H,2018MNRAS.476.4877M,2018MNRAS.474..547K,2020MNRAS.497.2265S}), the distribution of dark and baryonic matter in the cosmic web (e.g. \citealt{1999ApJ...514....1C,2001ApJ...552..473D,2014MNRAS.441.2923C,2017ApJ...838...21Z,2019MNRAS.486.3766M}). Filaments mark the transition from the low-density to the high-density environment and contain the largest portion of matter with respect to other structures, i.e., voids, wall/sheets, nodes/clusters at $z \lesssim 2$. The cosmic filaments could be used to probe the cosmology models and nature of gravity (e.g. \citealt{2018MNRAS.479..973C,2018A&A...619A.122H}), and may have played important roles in shaping the properties of halos and galaxies(e.g., \citealt{2007MNRAS.375..489H,2017MNRAS.466.1880C,2017A&A...600L...6K,2018MNRAS.474..547K}), in hosting the missing baryons (e.g. \citealt{1999ApJ...514....1C,2012ApJ...759...23S,2019A&A...624A..48D,2019MNRAS.483..223T}), etc. 

To reveal the roles of filaments in relevant studies, a comprehensive understanding of their properties is essential. This includes factors such as the length, density and velocity profiles of matter and galaxies embedded within them, as well as the gas temperature and local width/thickness. Once the filaments are identified with various methods, it is relatively straightforward to measure their length, as well as the galaxy overdensity profiles in their cross-sections for both the simulation and observation samples (e.g.\citealt{2005MNRAS.359..272C};\citealt{2010MNRAS.408.2163A}). However, measuring certain properties, such as width, is relatively difficult for a couple of reasons. Firstly, there is no unified definition of the filament boundary. This complexity is further compounded by the fact that galaxies serve as biased tracers of matter distribution. Secondly, filament width can vary considerably along their spine. Additionally, some properties, like the profiles of matter distribution and gas temperature, remain infeasible to be measured directly in filaments identified from observational data, despite having been extensively examined in simulation samples. (\citealt{2014MNRAS.441.2923C}; \citealt{2015MNRAS.453.1164G}; \citealt{2019MNRAS.486..981G}; \citealt{2021A&A...649A.117G}; \citealt{2020arXiv201209203T}; \citealt{2021ApJ...920....2Z};\citealt{2023arXiv230603966L}). On the other hand, the measured properties exhibit significant discrepancies attributed to variations in filament samples and the methodologies employed for their classification and measurement.

A lack of reliable estimation of filament properties will undoubtedly impede their roles in crucial issues such as galaxy evolution and the missing baryons at redshift $z<2$. For instance, regarding the latter issue, baryonic matter constitutes $\sim 5 \%$ of the energy density in the universe according to the standard $\Lambda \rm{CDM}$ cosmology (e.g., \citealt{2014A&A...571A..16P}). At redshift $z<2$, however, a significant portion, estimated to be between $30-50\%$, are `missing' from detection (\citealt{1998ApJ...503..518F};\ \citealt{2012ApJ...759...23S}; \citealt{2016ApJ...817..111D}). Most of those `missing' baryons are believed to be located in cosmic filaments and sheets ( \citealt{1999ApJ...514....1C};\ \citealt{2001ApJ...552..473D};\ \citealt{2006MNRAS.370..656D};\ \citealt{2016MNRAS.457.3024H}; \ \citealt{2017ApJ...838...21Z}; \ \citealt{2018MNRAS.473...68C}; \ \citealt{2019MNRAS.486.3766M}). X-ray emission and absorption of baryons in filaments have been reported in the literature (e.g., \citealt{2007ARA&A..45..221B};\ \citealt{2009ApJ...699.1765B}; \ \citealt{2015Natur.528..105E}; \citealt{2017A&A...606A...1A}; \citealt{2002ApJ...572L.127F};\ \citealt{2005ApJ...629..700N};\ \citealt{2016MNRAS.457.4236B};\ \citealt{2018Natur.558..406N};\ \citealt{2019A&A...621A..88N}; \ \citealt{2020A&A...643L...2T}). In addition, detection of the thermal Sunyaev-Zel'dovich (SZ) signal from baryons in filaments has recently been reported at 3-4 $\sigma$ level (\citealt{2018A&A...609A..49B};\ \citealt{2019MNRAS.483..223T};\ \citealt{2019A&A...624A..48D}; \ \citealt{2020A&A...637A..41T}). 

Future observations will enhance the statistical significance of detecting baryons within filaments through X-rays and the SZ effect. Despite this, accurately estimating the mass of baryons in filaments will remain a challenge, as certain properties, such as width, density, and temperature profiles, cannot be directly measured through observations. Currently, the width is typically assumed to be a fixed value, while other properties, such as the density and temperature distribution of baryons, rely on simulated results when estimating baryon content within filaments. However, there are notable discrepancies in the measured properties of filaments in simulation due to the differences in simulations and cosmic web classification. More studies and tools are needed to effectively reduce these uncertainties. On the other hand, properties of filaments measured in simulation should be further verified by observations, which urges the methods used to measure properties should be friendly for both simulation and observation samples. 

For example, measuring the width of filaments remains a challenging task. A number of works have estimated the radius of filaments based on the density profiles or the radial distribution of halos/galaxies derived from simulation or observation (\citealt{2005MNRAS.359..272C,2010MNRAS.408.2163A, 2010MNRAS.409..156B, 2010MNRAS.407.1449G, 2014MNRAS.438.3465T, 2016A&C....16...17T,2020A&A...638A..75B,2023MNRAS.525.4079Z}). However, distinctive results on the typical thickness have been reported, spanning 0.5-8.4 Mpc, which should be partly owed to the differences in filament samples and boundary definition. The radius inferred from matter distribution is expected to offer a more intrinsic measurement. \citealt{2005MNRAS.359..272C} found that the radial density profiles of filaments detected in simulations begin to follow a $r^{-2}$ power law at a radius of approximately 1.5-2.0 Mpc$/h$. \citealt{2010MNRAS.408.2163A} shows that the slope of the average filament profile in their simulation shifts from $-1$ to $-2$ at about r=2.0 $\rm{Mpc}/h$, which could be considered as the width of filaments. \cite{2014MNRAS.441.2923C} proposed measuring the local width of filaments in simulations based on the volume occupied by filament segments along the spine. Nonetheless, this method is difficult to apply directly to observational data.  \cite{2015MNRAS.453.1164G} quantified the total volume of filaments in simulations, allowing them to calculate the mean thickness under the assumption of cylindrical symmetry. However, the width of a single filament can vary significantly along the spine. 

These difficulties motivate us to seek a method or relation that can be used to estimate the local width of filaments constructed from simulations and observations. Moreover, the boundary of filaments that is defined by the matter distribution is favored. If there exists a correlation between the local width and the mass of dark matter halos enclosed in filament segments, it could be suitable. \cite{2014MNRAS.441.2923C} showed that the local width of filaments scales with the linear density, i.e., the mass per unit length along the spine. Hence, the key question is whether the linear density correlates with the mass contained by halos in filaments. In fact, halos form first in the denser region of the density field. Simultaneously, density peaks are linked by filaments. Consequently, denser filaments are anticipated to contain more massive halos. The findings of previous studies indicate that the total mass of matter in filaments is highly likely to correlate with the mass of halos embedded within. Filaments contain around $30\%-45\%$ of the total mass in the universe since $z=2$ (\citealt{2014MNRAS.441.2923C,2017ApJ...838...21Z,2018MNRAS.473.1195L, 2019MNRAS.486.3766M}).
Around $40\%$ of mass in the universe has been collapsed into halos at $z=0$, and this proportion decreases gradually with increasing redshift (e.g., \citealt{2016MNRAS.457.3024H}). Meanwhile, most of the dark matter halos more massive than $\sim 10^{10.5} \rm{M_{\odot}}$ are located in the filaments and knots since $z\sim 4$(\citealt{2014MNRAS.441.2923C,2022ApJ...924..132Z}).
\cite{2019MNRAS.486..981G} showed that the total mass of the filaments scales with the total mass of resident halos.  

Therefore, the key remaining question is whether there is a correlation between the mass of filaments and the embedded halos per unit length or not. In this paper, we will investigate this aspect and further probe the relation between the local width of cosmic filaments and the mass of dark matter halos enclosed, utilizing samples from a cosmological hydrodynamic simulation. This paper is organized as follows. Section 2 provides a brief description of the cosmological simulation and the numerical methods we employed to identify filaments and measure their width, total mass, and mass of enclosed halos per unit length. Section 3 presents the relationships we find. We discuss future improvements and potential applications in Section 4. Finally, our findings are summarized in section 5.   

\section{Methodology} \label{sec:method}

\begin{figure}[htbp]
\begin{center}
\hspace{-0.0cm}
\includegraphics[width=0.65\textwidth, trim=350 1350 20 220, clip]
{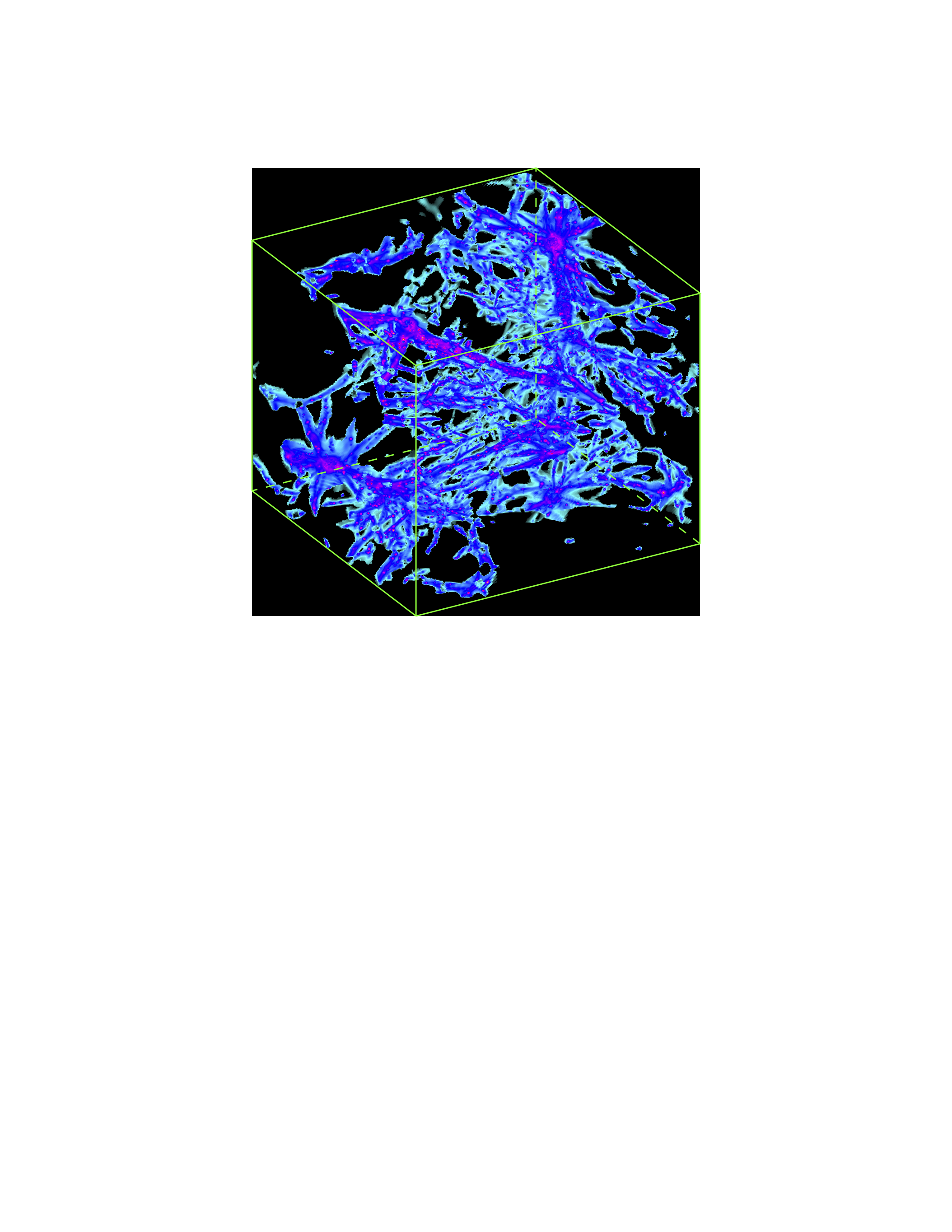}
\includegraphics[width=0.55\textwidth, trim=20 1080 20 100, clip]{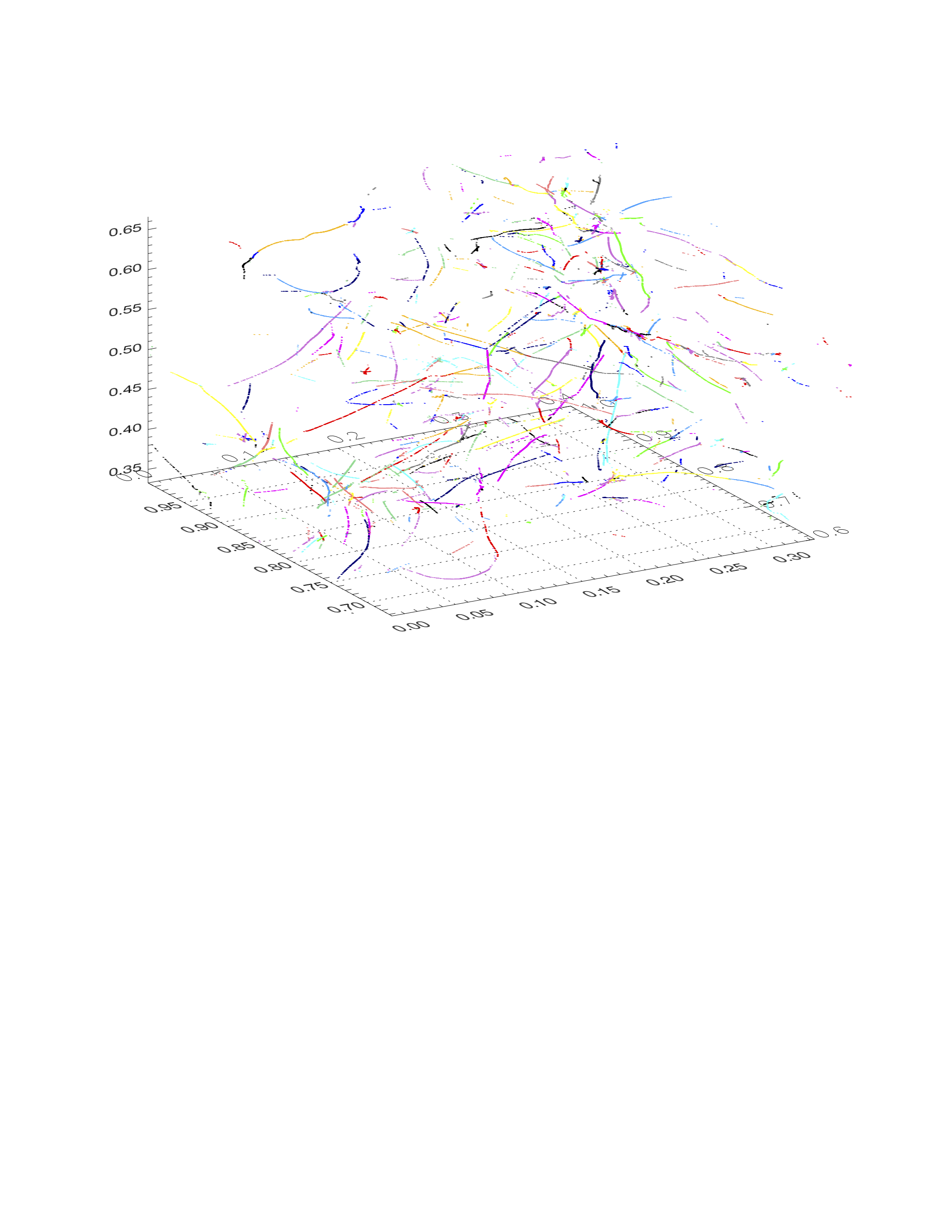}
\includegraphics[width=0.50\textwidth, trim=0 350 60 100, clip]{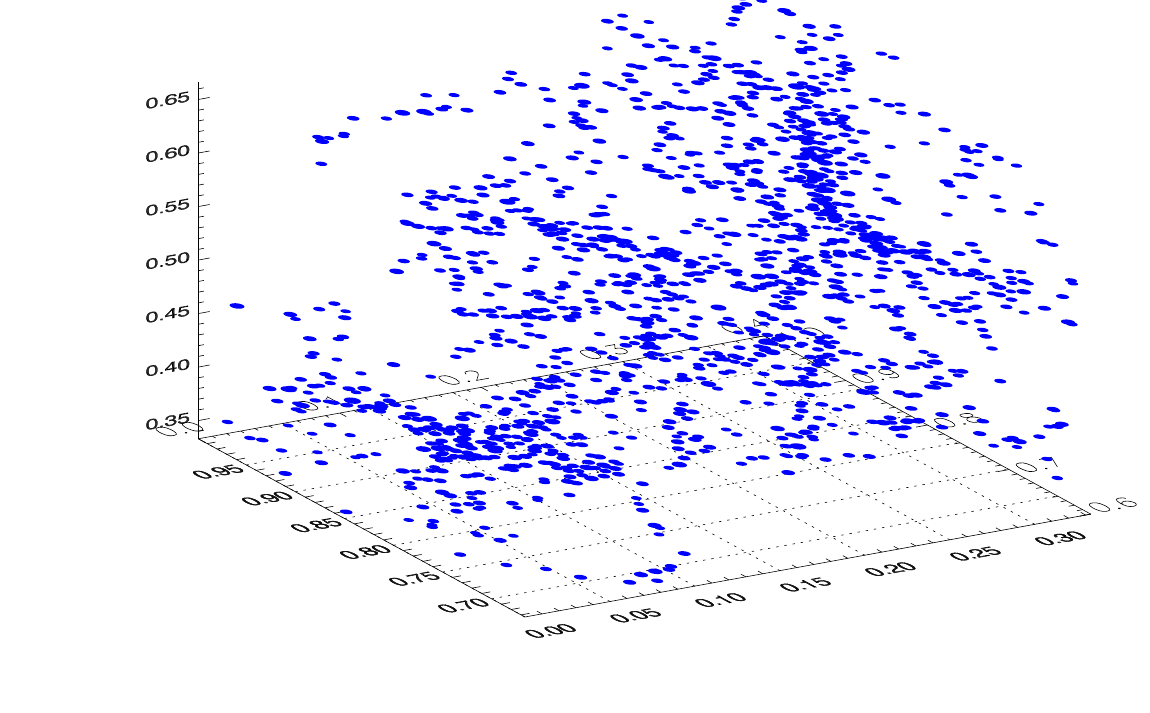}
\caption{Top: Density field of baryonic matter belonging to filaments and nodes in a cubic box of $(33.3 h^{-1}\rm{Mpc})^3$; Middle: $1\%$ of the grids belonging to filaments after compressing. Grids in the same filament segment are coded with the same color. Due to the limited color slots, different segments may share the same color. Bottom: Dark matter halos in filaments and nodes.}
\label{fig:fila_segment}
\end{center}
\end{figure}

\subsection{Simulation } 
Our cosmological hydrodynamic simulation sample was produced by the code RAMSES(\citealt{2002A&A...385..337T}), adopting a $\Lambda$CDM cosmology ($\Omega_{m}=0.317, \Omega_{\Lambda}=0.683,h=0.671,\sigma_{8}=0.834, \Omega_{b}=0.049$, and $n_{s}=0.962$, \citealt{2014A&A...571A..16P}). The simulation traces the evolution of matter from $z=99$ to $z=0$ in a volume of $(100 h^{-1})^3$ Mpc. It utilizes  $1024^3$ dark matter particles and $1024^3$ root grid cells, i.e. each with a side length of $97.6 h^{-1}$kpc. Adaptive mesh refinement is employed, achieving a resolution of $0.763 h^{-1}$kpc at the finest level. At redshift $8.5$, a uniform UV background is turned on following the model in \cite{1996ApJ...461...20H}. Various sub-grid physical modules, such as gas cooling and heating, star formation, and stellar feedback, are implemented, excluding feedback from active galactic nuclei(AGN). We have identified the dark matter halos in our simulation samples using the Friend-of-Friend (FOF) algorithm. For more details on the simulation, readers can refer to \cite{2021ApJ...906...95Z}.  

\subsection{Filaments processing and measurements}

The cosmic web in our samples, including structures such as nodes, filaments, sheets, and voids, is identified using the tidal tensor of the rescaled peculiar gravitational potential, which is usually named T-Web and is one of the widely used algorithms. This method can capture the dynamical evolution of cosmic structures by analyzing the tidal force field. We have used the same procedure and parameters selection as in \cite{2021ApJ...920....2Z} for classifying the cosmic web and compressing the filaments. The threshold eigenvalue $\lambda_{th}$ is an important parameter in the T-Web algorithm regarding the properties of identified cosmic structures. We have utilized a default value of $\lambda_{th}=0.2$, which can provide a better visual impression of the web (\citealt{2009MNRAS.396.1815F, 2017ApJ...838...21Z}). Moreover, the average density at the boundary of the identified filaments with $\lambda_{th}=0.2$ is around 1.8 times the cosmic mean(\citealt{2021ApJ...920....2Z}), which can be considered a good separation between the filaments and walls. Subsequently, we adopt the methods in \cite{2014MNRAS.441.2923C} for the compression of filaments. More details about the construction of T-Web and filament compression can be found in \cite{2021ApJ...920....2Z}. Note that all grid cells in filaments and nodes are included when we compress the filaments and measure their properties. On the other hand, we have used an alternative approach, instead of the method in \cite{2014MNRAS.441.2923C}, to separate the filaments into segments, as detailed below.

\begin{enumerate}
\item  
Starting from any one grid, indicated as $g_i$ where i is the tag number, within the compressed filaments, we firstly search its neighboring grids within a sphere of small radius $r_i=0.04\, \rm{Mpc/}h$, which is roughly 0.2 times of the grid size used for constructing the T-Web. Subsequently, we calculate the angles between the orientation at different neighboring grids and that at grid $g_i$. If the angle for a neighboring grid is less than $\theta_{th}=30$ degrees, it is categorized as part of the same filament segment as grid $g_i$. Grids in the same segment will be allocated the same segment number. The rationale behind choosing $\theta_{th}=30$ degree is as follows: The visual impression shows that a $\theta_{th}$ smaller than 20 degrees struggles to handle slightly curved segments, while a $\theta_{th}$ larger than 40 degrees fails to segment the filament network into distinct branches.
\item 
The procedures in Step 1 are executed for each grid within the filaments, following which all the grids are assigned to one of the filament segments. We store the information for each segment in the list.  
\item 
The filament segments identified in the preceding two steps are relatively short and are called primary segments. Hence, it is necessary to consolidate nearby short primary segments. For each primary segment, we walk along its spine to search for all neighboring segments within a distance of $r_{neb}$. The closest pair of grids across a primary segment and any neighboring primary segments is determined, i.e., with one grid from each of the segment pairs. If the angle between the orientations at this pair of grids is less than $\theta_{th}=30$ degrees, their host primary segments will be linked to form a longer segment. 
\item 
Step 3 is iterated multiple times, with the search radius $r_{neb}$ gradually increasing from $0.04\, \rm{Mpc/}h$ to $0.4\, \rm{Mpc/}h$. We have assessed the performance of our method with various maximum $r_{neb}$ values, denoted as $r_{neb,max}$, ranging from 0.2 to 1.0 Mpc$/h$. We find that a $r_{neb,max}$ lower than 0.3 Mpc$/h$ leads to excessively fragmented segments, whereas a $r_{neb,max}$ greater than 0.5 Mpc$/h$ tends to merge relatively isolated segments via bridges. Therefore, we select an $r_{neb,max}=0.4$ Mpc$/h$ as a balanced choice.
\end{enumerate}

A total of 54514 segments have been identified in the sample at redshift z=0. The upper panel in Figure \ref{fig:fila_segment} illustrates the density of grid cells within filaments and nodes in a cubic box with a side length of $33.3 \rm{Mpc/}h$ at $z=0$. The middle panel of Figure \ref{fig:fila_segment} displays the filament segments after compression and segmentation within the same box. Grid cells within the same segment are color-coded identically. Note that, due to a limited number of colors in the color table, filament segments in different regions may share the same color. The visual assessment suggests that our segmentation procedure works reasonably well. The lower panel of Figure \ref{fig:fila_segment} shows the distribution of dark matter halos within the same cubic volume. We can see that the distribution of halos mirrors the frame of the filaments. 

After segmenting the filaments, we measured the length, local width/diameter, $D_{\rm{fil}}$, and liner mass density of filament segments, $\zeta_{fil}$, following the procedures introduced in \cite{2014MNRAS.441.2923C}. At each grid within any filament segment, $g_i$, we place a sphere with a radius of $R_f=1.5 \rm{Mpc/}h$. This value is moderately lower than the value used in \cite{2014MNRAS.441.2923C} and \citealt{2021ApJ...920....2Z}), because it can more effectively capture the variations of local width. The number of grids encompassed within the sphere is counted, denoted as $N_{gs}$. Subsequently, the contribution of grid $g_i$ to the filament length is calculated as $\delta L = 2*R_f/N_{gs}$. The local thickness/width of filaments at grid $g_i$ is determined as $D_{fil} = \sqrt{4V_{voxel}N_{gs}/(2\pi\,R_f)}$, where $v_{voxel}$ represents the volume of a grid cell. The total mass contained by the grids within the sphere, $\Delta M$, is utilized to compute the linear mass density of the filament segment at grid $g_i$, denoted as $\zeta_{fil}=\Delta M/(2R_f)$. When estimating these filament properties, only grid cells within the same filament segment are accounted for by default. This restriction will be relaxed in some cases to demonstrate its impact. Moreover, we link each of the dark matter halos within the filaments and nodes to the grid cell that hosts the halo center and subsequently to the parent segment containing this grid cell, which is called the primary host segment of a halo. It is important to note that halos more massive than $10^{12} \rm{M_\odot}$ are connected by multiple filaments, with a typical number of 2-3 (e.g., \citealt{2010ApJ...723..364A,2018MNRAS.473...68C}). To take this fact into account, we also assign halos more massive than $10^{12} \rm{M_\odot}$ to a secondary parent segment, specifically the one that contains the grid with the shortest distance to the halo center among those grids outside of the primary parent segment.

\begin{figure}[htbp]
\begin{center}
\hspace{-0.0cm}
\includegraphics[width=0.55\textwidth, trim=60 360 20 80, clip]{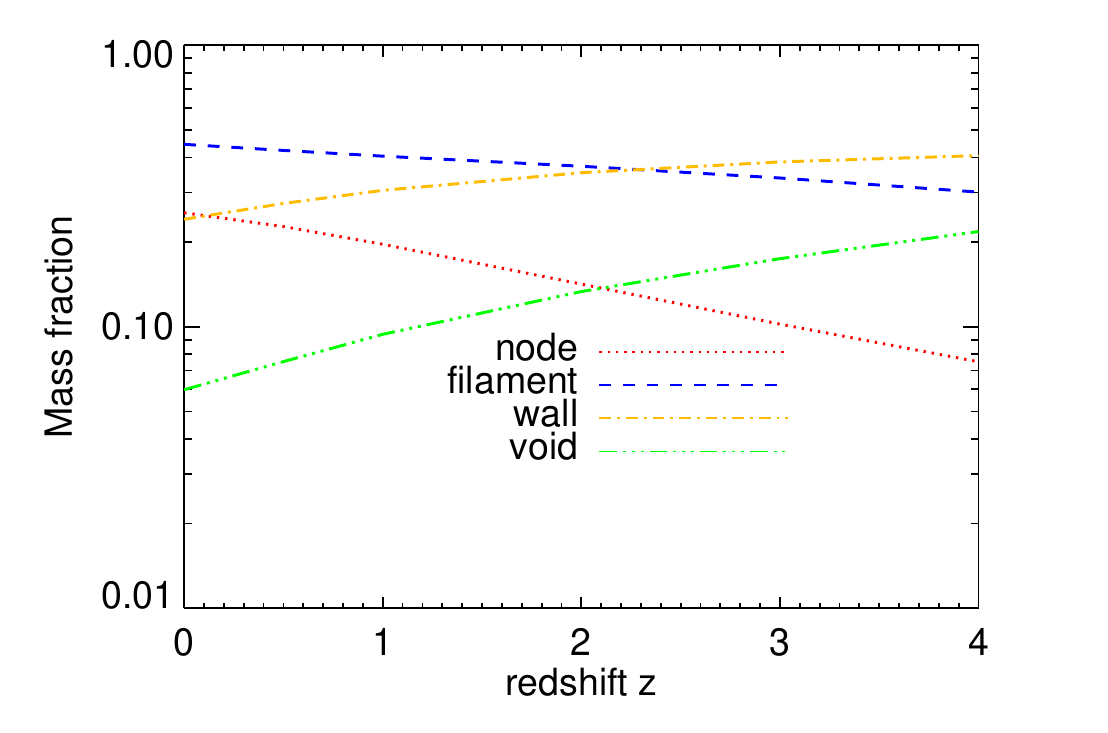}
\includegraphics[width=0.55\textwidth, trim=60 360 20 80, clip]{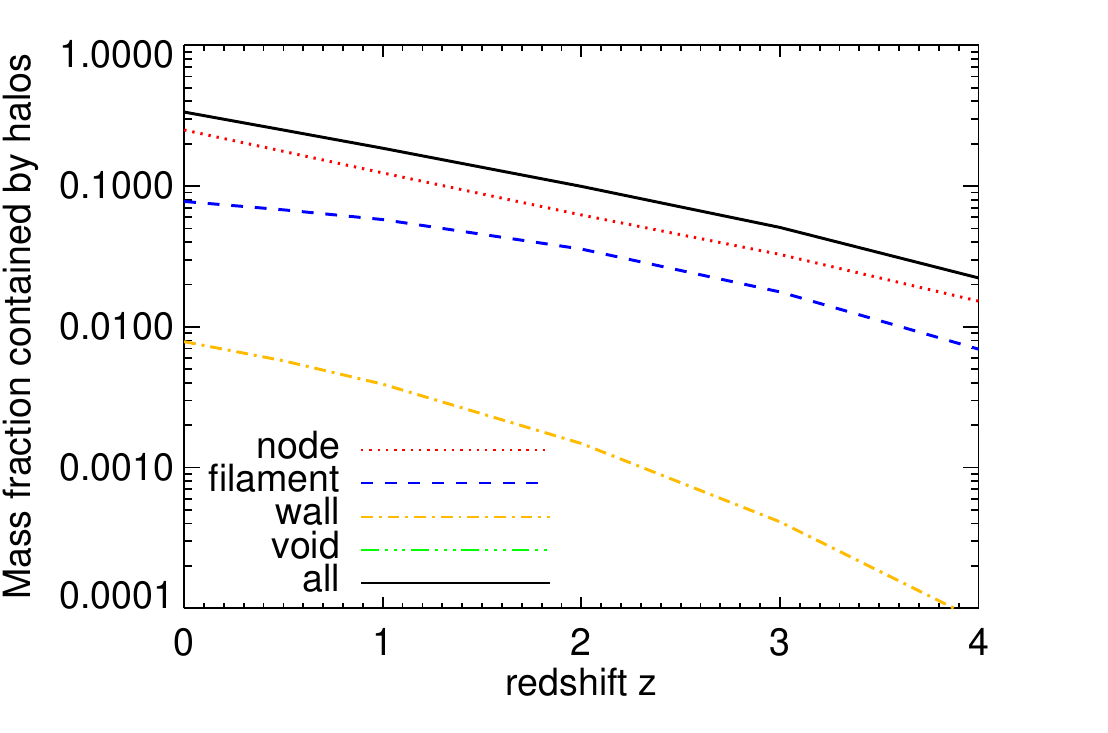}
\caption{Top: Fractions of mass residing in various structures from redshift $z=4$ to $z=0$. Bottom: Fractions of mass contained by halos that reside in different structures.}
\label{fig:mass_frac}
\end{center}
\end{figure}

\subsection{Linear halo mass density}
Given the correlation between local width and linear density of filaments, if there exists another measurement that scales with linear density, it is expected to exhibit a correlation with local width. One such a candidate is the mass of halos within filaments per unit length. Cosmic filaments and nodes are structures that have undergone collapse in at least two of the three dimensions, whereas halos have achieved virial equilibrium after three dimensions collapse. Consequently, the proportion of the mass in filaments and nodes occupied by embedded halos is anticipated to surpass the overall fraction of the mass held by halos in the whole universe. In our sample, approximately $32\%$ of the mass in the universe is contained by the halos in filaments and nodes. Meanwhile, filaments and nodes contain around $70\%$ of cosmic mass. Namely, around $46\%$ of the mass in the filaments and nodes are contained by halos. The evolution of these fractions as a function of redshift is depicted in Figure \ref{fig:mass_frac}. 

If we divide filament segments into numerous short slices along their spines and calculate the mass of matter in embedded halos within those short slices, the results would exhibit considerable fluctuations. However, when measured in slices of length 1 Mpc/h or longer, these fluctuations would be reduced. Therefore, we introduce a novel measurement, the linear halo mass density of filaments, $\eta_{fil}$, representing the total mass of halos embedded in filaments per Mpc/h.
The measurement of $\eta_{fil}$ is akin to the linear mass density of filaments but with some differences. At any grid cell along the filament spine, denoted as $g_j$, we set up a virtual sphere centered at $g_j$ with a radius of $R_f=1.5 \rm{Mpc/}h$. We identify all the dark matter halos whose centers are within this sphere. Subsequently, we select halos hosted by the same filament segment as grid $g_j$, including both primary and secondary hosts, in the default mode. The FOF mass of those selected halos is summed up and defined as $\Delta M_h$. Consequently, $\eta_{fil}$ at grid $g_j$ is defined as $\eta_{fil}=\Delta M_h/(2R_f)$. Note that we will relax the limitation that only halos in the same filament segment are included when calculating $\Delta M_h$ and $\eta_{fil}$ if the similar limitation is lifted when calculating $\zeta_{fil}$. The scaling relations between the linear halo mass density, linear mass density, and the local width of cosmic filaments will be explored in the subsequent section. 

\begin{figure}[htbp]
\begin{center}
\hspace{-0.0cm}
\includegraphics[width=0.5\textwidth, trim=60 360 20 80, clip]{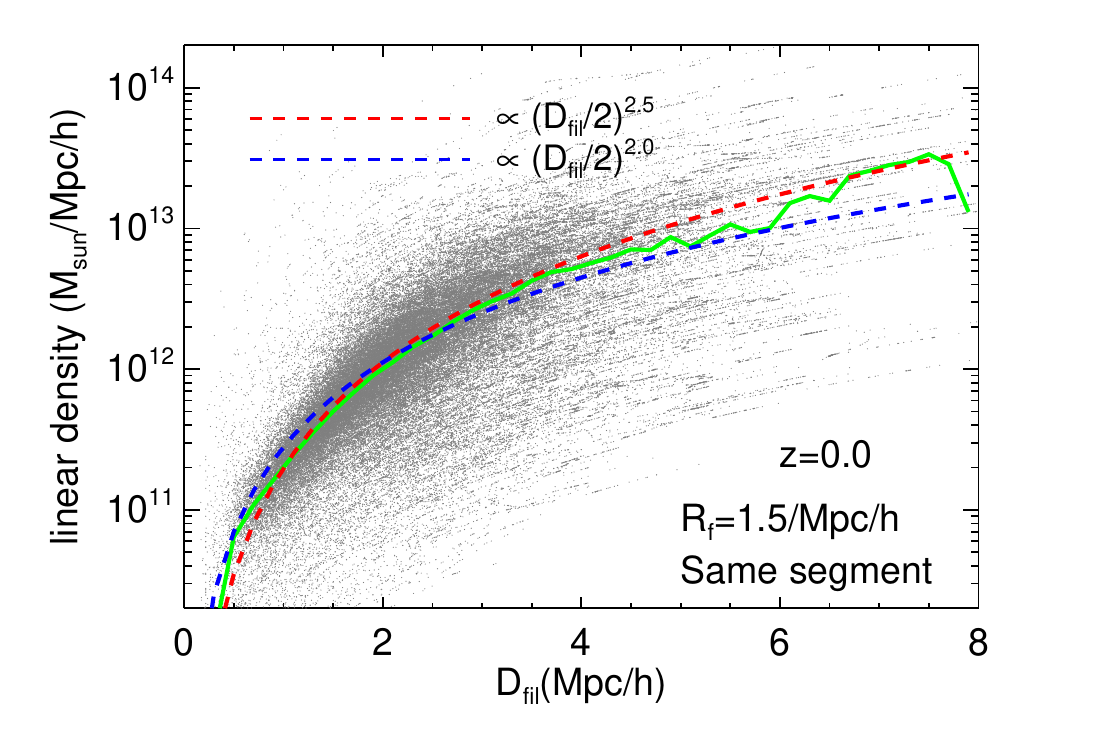}
\includegraphics[width=0.5\textwidth, trim=60 360 20 80, clip]{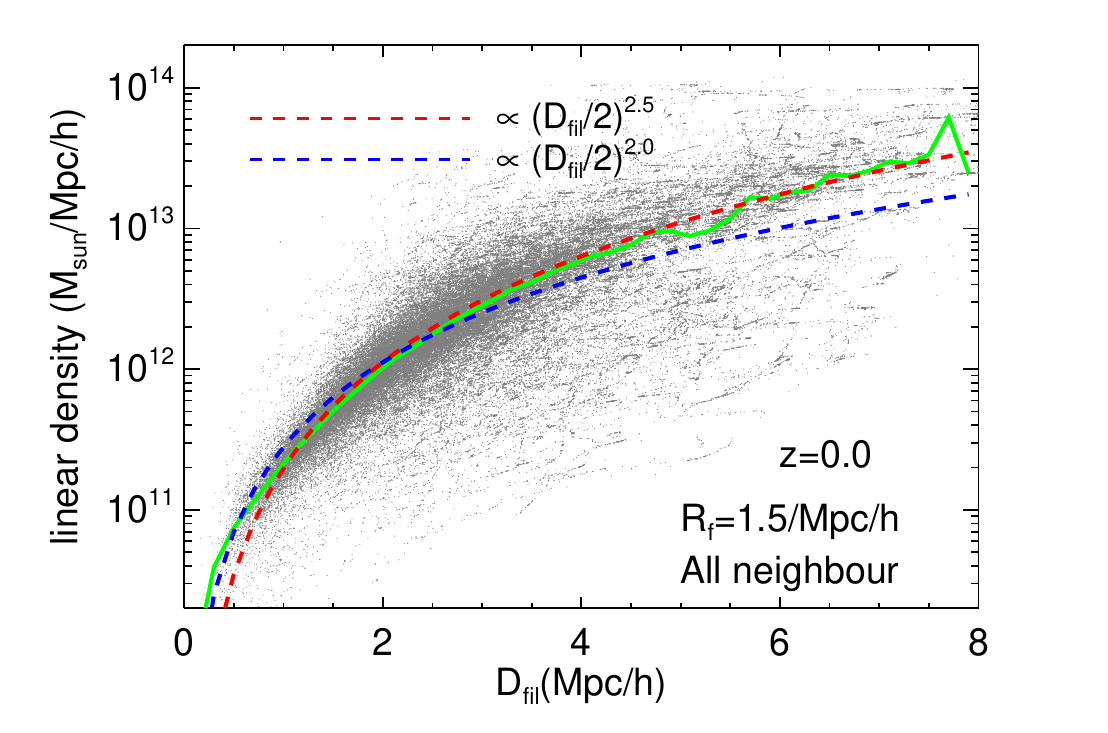}
\caption{Top: The local thickness versus the linear density that is estimated at grids in filaments, requiring that the neighboring grids used to estimate the filament properties are in the same segment. The solid green line indicates the median linear density. Bottom: The same as the left, but all the neighboring grids in filaments are accounted for when estimating the linear density and local thickness.}
\label{fig:dia_linmass}
\end{center}
\end{figure}

\section{Local width-linear halo mass density relation} \label{sec:fila_width}
\subsection{linear mass density - linear halo mass density relation}

\begin{figure*}[htb]
\begin{center}
\vspace{-0.0cm}
\includegraphics[width=0.45\textwidth, trim=60 360 20 80, clip]{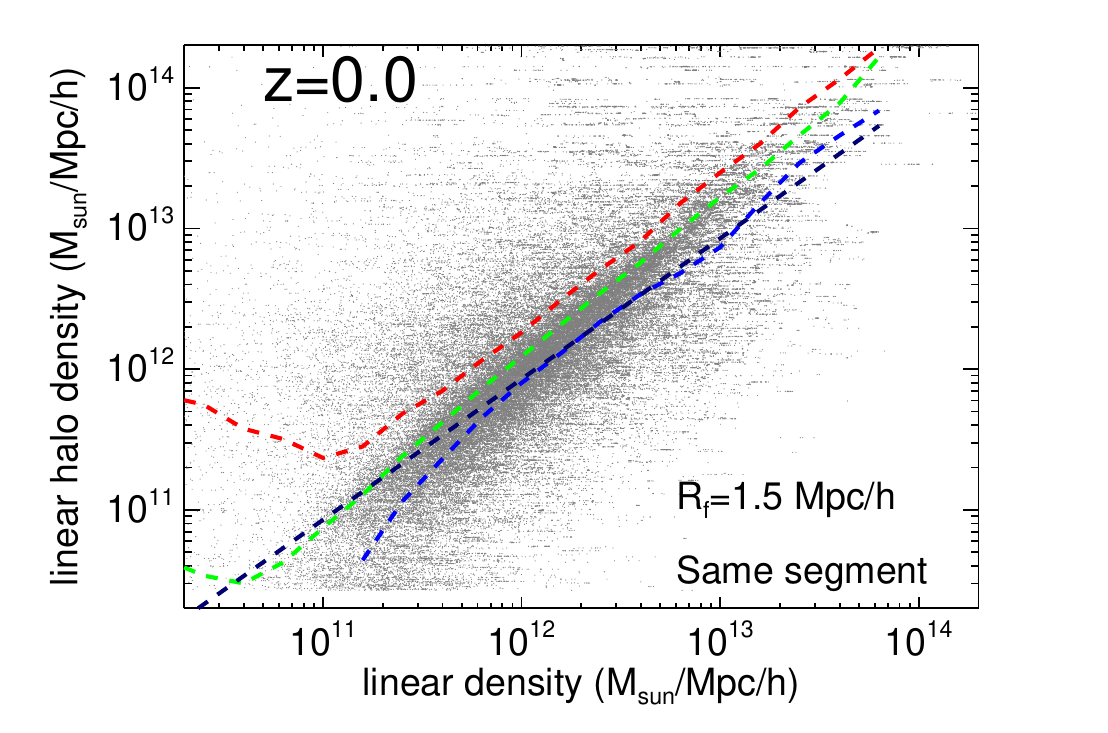}
\includegraphics[width=0.45\textwidth, trim=60 360 20 80, clip]{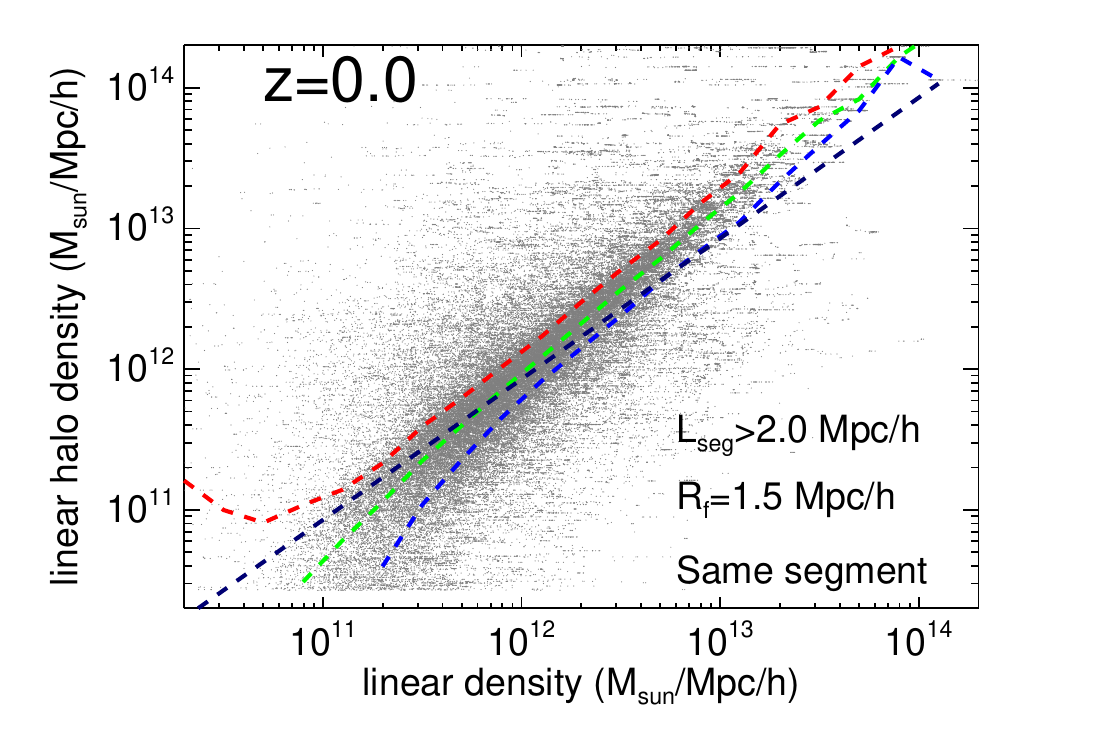}
\includegraphics[width=0.45\textwidth, trim=60 360 20 80, clip]{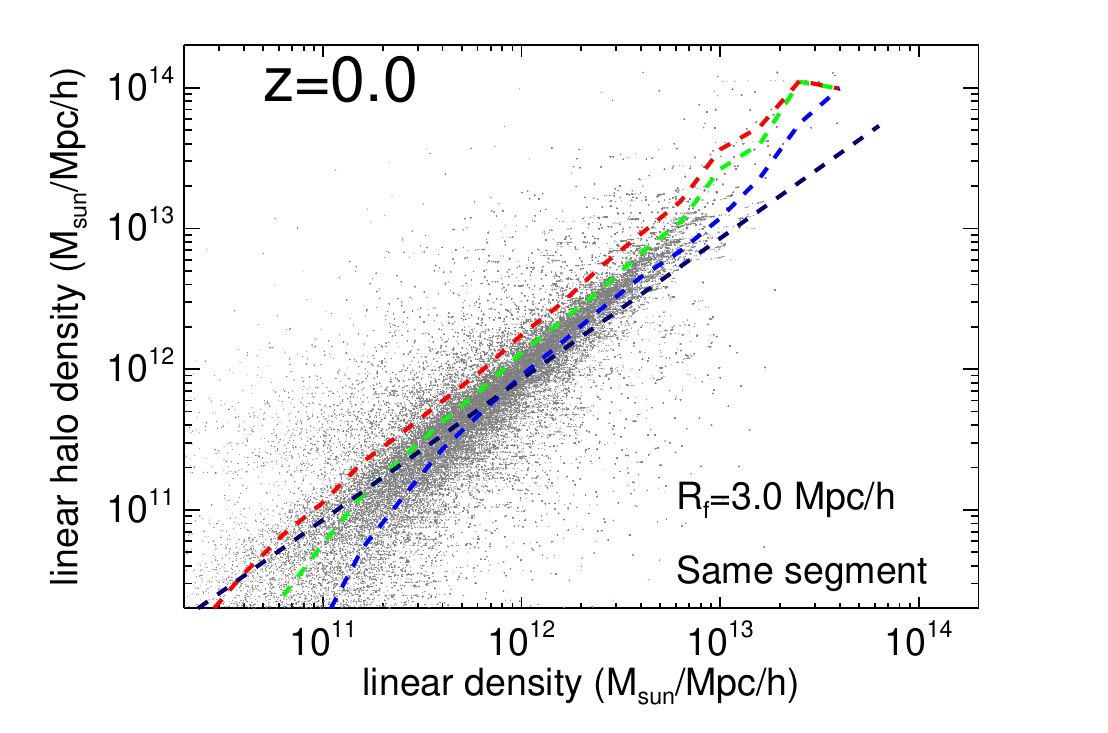}
\includegraphics[width=0.45\textwidth, trim=60 360 20 80, clip]{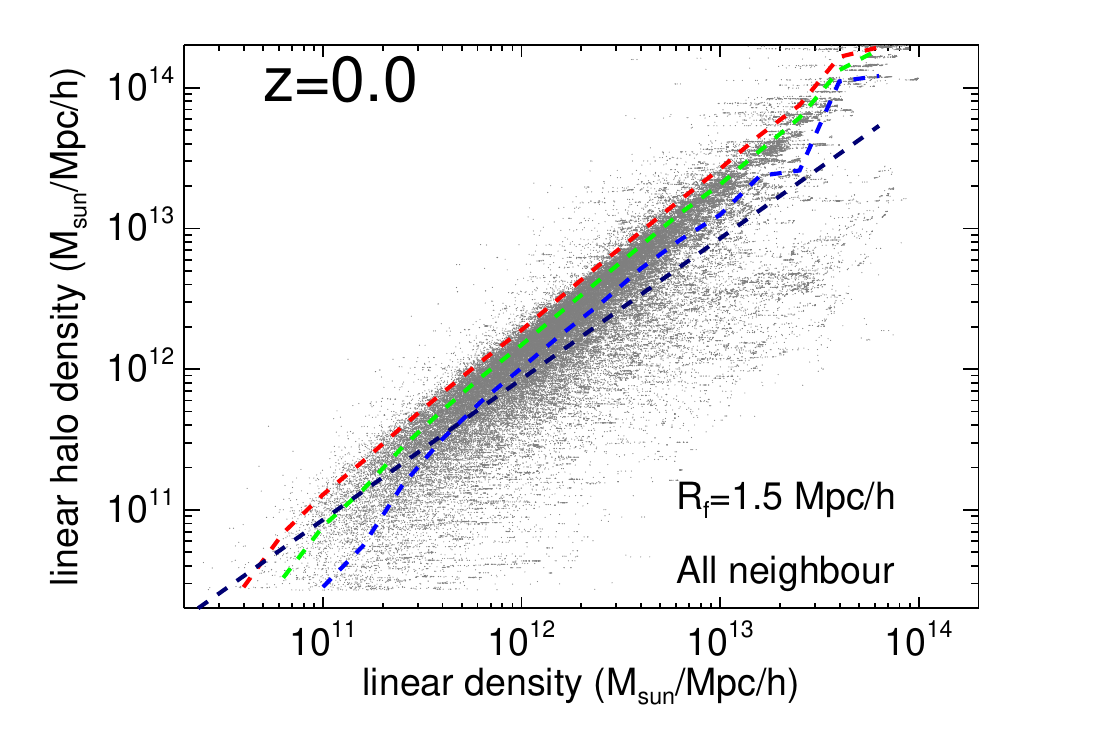}
\caption{Top left: The correlation between linear density of filaments and linear halo mass density that measured at 150,000 grids in filaments. Grid cells in this plot have been randomly selected from all the grids in filaments and account for around $7\%$ of the whole sample. The red, green, and blue dashed lines indicate the 75, 50, and 25 percentile, respectively, in each linear density bin. Black line shows $\eta_{fil}=0.85 \zeta_{fil}$. Top right: The same as top left, but select grids from filament segments longer than $2.0\, \rm{Mpc}/h$. Bottom left: The same as top left, but with a neighboring search radius $R_f=3.0\, \rm{Mpc}/h$. Bottom right: The same as the top left, but all the neighboring grids and halos in filaments are accounted for when estimating the linear density and linear halo density. }
\label{fig:fila_linden_halo}
\end{center}
\end{figure*}

\cite{2014MNRAS.441.2923C} found a scaling relation between the local width/diameter of filaments and the local linear mass density, $\zeta_{fil}$, in N-body simulation samples, albeit exhibiting considerable scatter. Based on the cosmological hydrodynamical simulation, we (\citealt{2021ApJ...920....2Z}) demonstrated that the local width of filaments is indeed correlated with the local linear mass density, aligning with \citealt{2014MNRAS.441.2923C}. In \cite{2021ApJ...920....2Z}, where the filament segmentation process is not applied, we found that the median of $\zeta_{fil}$ as a function of local diameter can be approximately described as

\begin{equation}
    \zeta_{fil} \approx  \frac{ 2.7\times 10^{11}\rm{M}_{\odot}}{(\rm{Mpc}/h_0)} \times \pi\times(\frac{R_{fil}}{(\rm{Mpc}/h_0)})^n,
\label{eqn:fila_linden}
\end{equation}

where the power index n is approximately $2-2.5$. After the filament segmentation procedure, the correlation between the local thickness and the linear mass density is shown in the top panel of Figure \ref{fig:dia_linmass}. The solid green line represents the median linear density as a function of $D_{fil}$, which can still be fitted by eqn. \ref{eqn:fila_linden} with a power index n ranging from $2-2.5$. However, there is noticeable scatter in the $D_{fil}-\zeta_{fil}$ relation. Allowing for the inclusion of grids beyond the same segment when estimating $D_{fil}$ and $\zeta_{fil}$ will moderately reduce the scatter, as illustrated in the bottom panel of Figure \ref{fig:dia_linmass}. Note that our objective is to find a viable statistical estimator for the width of filament segments. Achieving high accuracy at each grid within the filaments is not an immediate necessity at this stage. Therefore, we will accept the scatter in the $D_{fil}-\zeta_{fil}$ relation in this study.

For the sample at redshift z=0, we have measured the linear mass density and the linear halo mass density at $\sim$20 millions grid cells, with a neighboring search radius of $R_f=1.5\, \rm{Mpc}/h$. 
The top left panel in Figure \ref{fig:fila_linden_halo} illustrates the results of $\eta_{fil}$ versus $\zeta_{fil}$ at 100,000 randomly selected grids from the entire sample. Observing Figure \ref{fig:fila_linden_halo}, one can discern an overall scaling relationship between these two characteristics. However, there is a noticeable scatter in the $\zeta_{fil}-\eta_{fil}$ correlation, particularly at the low and high linear mass density ends. The red, green, and blue dashed lines indicate the 75th, 50th, and 25th percentile of linear halo mass within each linear mass density bin for the entire sample. The median linear halo mass as a function of the linear mass could be roughly described by a linear relation as 

\begin{equation}
    \eta_{fil} \approx f_h*\zeta_{fil}
\end{equation}
, where $f_h \sim 0.85-0.90$. 

\begin{figure*}[htb]
\begin{center}
\hspace{-0.0cm}
\includegraphics[width=0.45\textwidth, trim=60 360 20 80, clip]{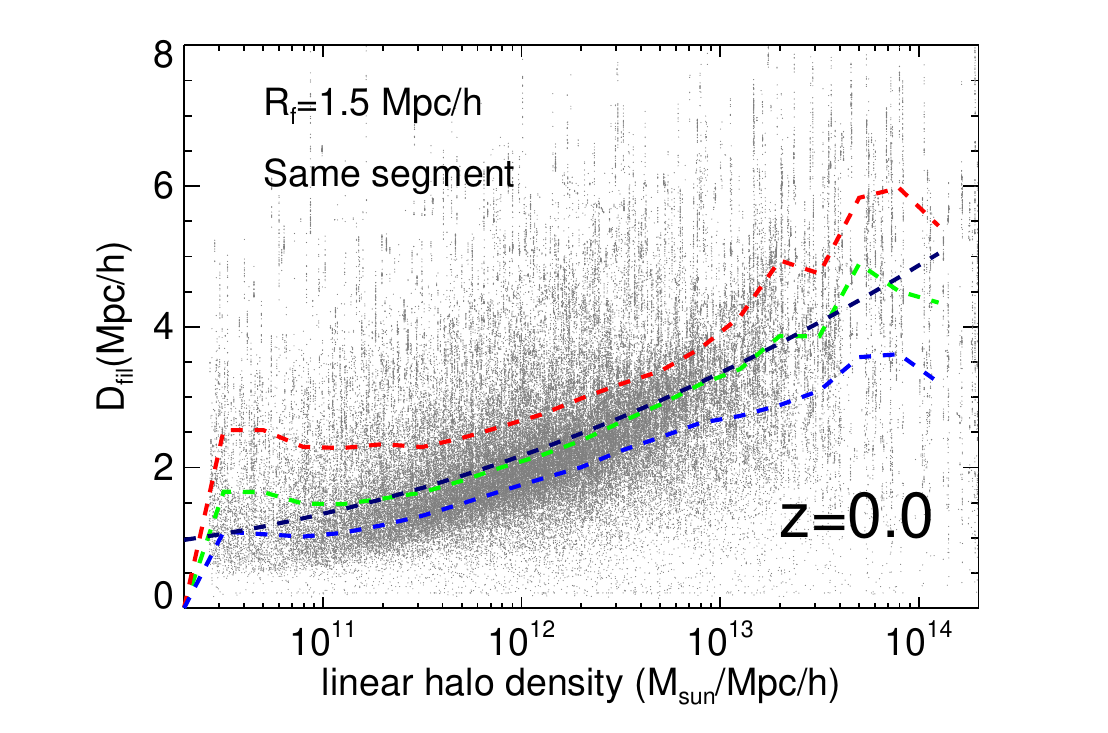}
\includegraphics[width=0.45\textwidth, trim=60 360 20 80, clip ]{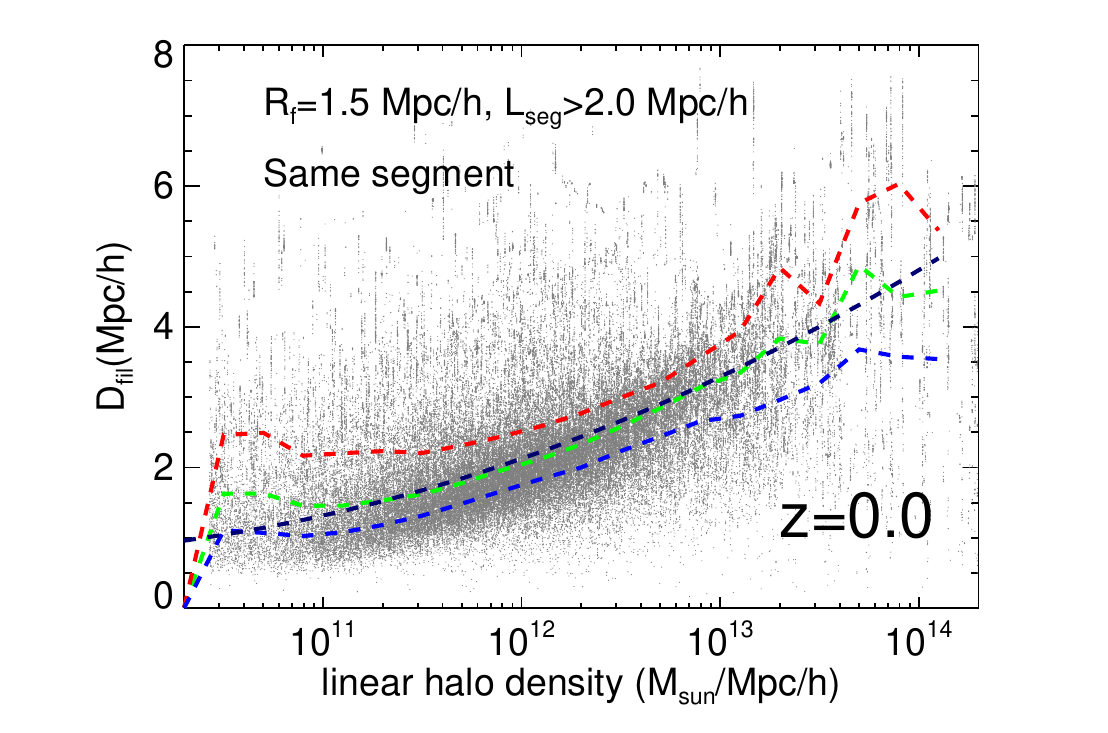}
\includegraphics[width=0.45\textwidth, trim=60 360 20 80, clip ]{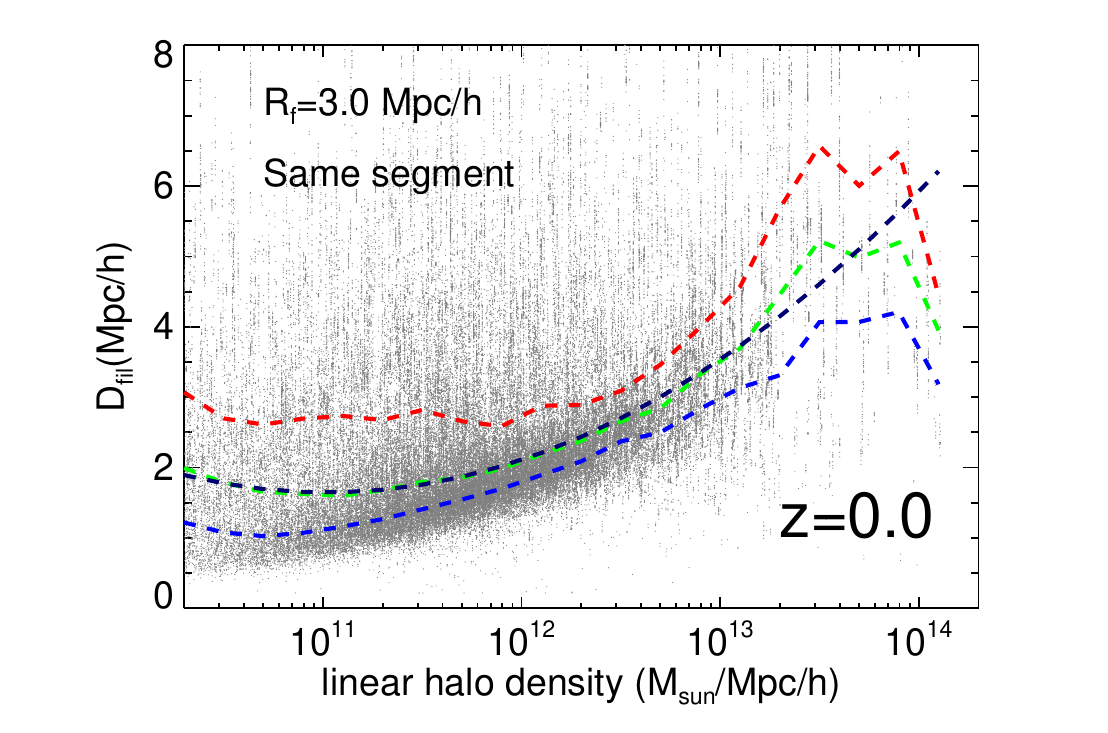}
\includegraphics[width=0.45\textwidth, trim=60 360 20 80, clip ]{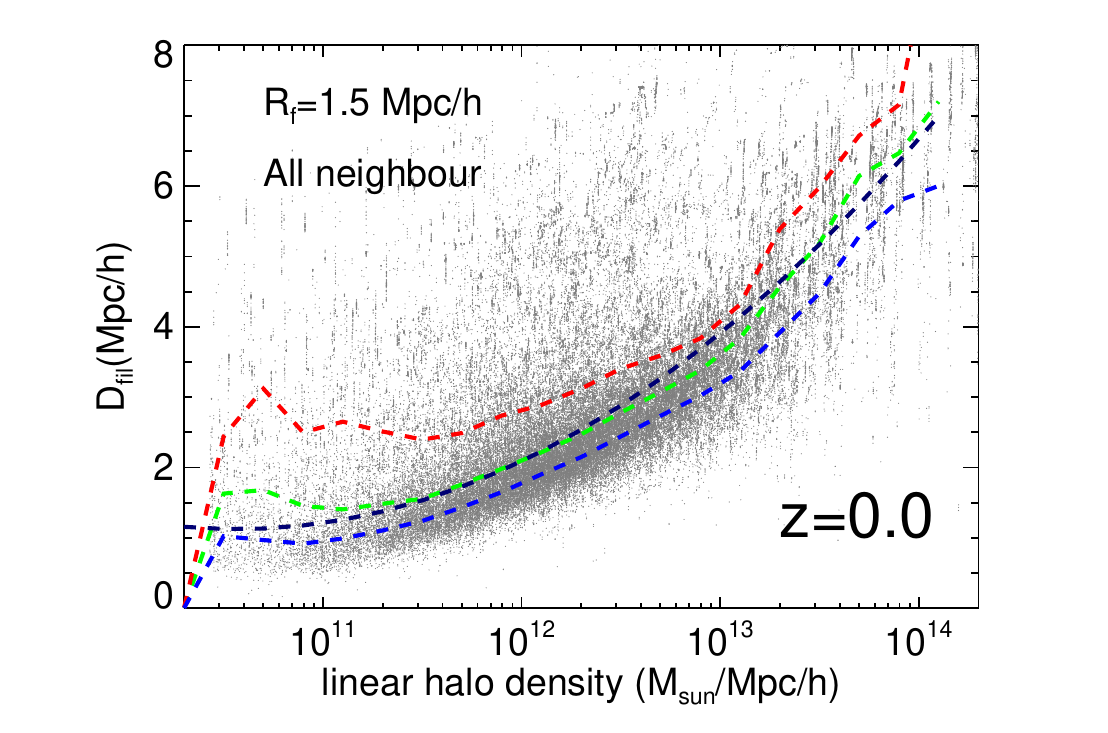}
\caption{Top Left: The linear halo density and local thickness estimated at 100,000 grids in filaments. The red, green, and blue dashed lines indicate the 75, 50, and 25 percentile, respectively, in each linear halo density bin. Top Right: Same with the left, but only for grids selected from filaments with lengths longer than 2 $\rm{Mpc}/h$. Bottom left: The same as top left, but with a neighboring search radius $R_f=3.0\, \rm{Mpc}/h$. Bottom right: The same as the top left, but all the neighboring grids and halos in filaments are accounted for when estimating the linear density and linear halo density. }
\label{fig:filb_halo_width}
\end{center}
\end{figure*}

This correlation aligns with our expectations based on the discussions in the preceding sections. However, $f_h$ significantly exceeds the overall mass fraction contained by halos in the filaments and nodes, as depicted in Figure \ref{fig:mass_frac}. This disparity primarily stems from the mean value of  $\eta_{fil}/\zeta_{fil}$ being lower than the median value, due to a substantial portion of grid cells having a quite small $\eta_{fil}$, as illustrated in Figure \ref{fig:fila_linden_halo}. Simultaneously, the challenge of accurately estimating $\eta_{fil}$ and $\zeta_{fil}$ for grids within short filament segments likely contributes to the scatter in the $\zeta_{fil}-\eta_{fil}$ relation. Grid cells within segments shorter than 2.0 $\rm{Mpc}/h$ constitute 18$\%$ of all grid cells in filaments and nodes. If these segments shorter than 2.0 $\rm{Mpc}/h$ are excluded, the scatter in the $\zeta_{fil}-\eta_{fil}$ relationship will be moderately reduced, illustrated in the top right panel of Figure \ref{fig:fila_linden_halo} shows. On the other hand, doubling the neighboring search radius $R_f$ used in estimating $\eta_{fil}$ and $\zeta_{fil}$ to 3.0 $\rm{Mpc}/h$ will significantly reduce the scatter in the $\zeta_{fil}-\eta_{fil}$ relation, as demonstrated in the bottom left panel of Figure \ref{fig:fila_linden_halo}. However, a large $R_f$ will wash out some of the local variations in the properties of the filament, as noted by \cite{2014MNRAS.441.2923C}. Alternatively, eliminating the constraints that only grids and halos in the same filament segments are considered when estimating $\eta_{fil}$ and $\zeta_{fil}$ would also lead to a significant reduction in the scatter of the $\eta_{fil}$-$\zeta_{fil}$ relationship, as shown by the bottom right panel of \ref{fig:fila_linden_halo}. 

\subsection{Linear halo mass - filament width relation}

We now move on to explore the relation between the linear halo mass density and the local width/thickness of grids in filaments. Figure \ref{fig:filb_halo_width} presents the results of $\eta_{fil}$ versus $D_{fil}$ across 100,000 grids. In the top left panel, grids are selected from all the filament segments, with a neighbor-finding radius $R_f=1.5\, \rm{Mpc}/h$ used for estimating $\eta_{fil}$ and $D_{fil}$. An overall correlation between these properties is observed, although there exists a noticeable scatter in the filament width at a given $\eta_{fil}$. To quantitatively assess this scatter, we plot the 75th, 50th, and 25th percentile filament widths for all the grids in each linear halo mass bin using red, green, and blue dashed lines. The interquartile range has a minimum value of approximately $1 \rm{Mpc}/h$ within the range $\eta_{fil}\sim 10^{12.2}-10^{13.2}\rm{M_\odot/Mpc}/h$, and gradually increasing towards both lower and higher $\eta_{fil}$ ends. We use a second-order polynomial to fit the median value of $\rm{D_{fil}}$ in each $\eta_{fil}$ bin. The fitting curve is represented by the black dashed line, which can be interpreted as 

\begin{equation}
\begin{aligned}
\frac{\rm{D_{fil}}}{\rm{Mpc}/h}=0.865+0.301\times log_{10}(\eta_{fil,10})+\\0.175\times [log_{10}(\eta_{fil,10})]^2, 
\label{eqn:dfil_eta_all}
\end{aligned}
\end{equation}

where $\eta_{fil,10}=\frac{\eta_{fil}}{10^{10}\, \rm{M_\odot/Mpc}/h}$. The corresponding $\rm{D_{fil}}$ values for $\eta_{fil}=10^{11},10^{12}$ and $10^{13}\,\rm{M_\odot/Mpc}/h$ predicted by eqn. \ref{eqn:dfil_eta_all} are $1.341, 2.167$ and $3.343 \rm{Mpc}/h$ respectively. On the other hand, $49.7\%$ of the grid cells in filaments lie within a range of $\pm 0.5\rm{Mpc}/h$ around the values given by eqn. \ref{eqn:dfil_eta_all}. 

The top right panel shows the case if only filament segments longer than 2 $\rm{Mpc}/h$ are taken into account, resulting in a moderate reduction in the scatter observed in the $\eta_{fil}$-$D_{fil}$ relation. Consequently, the fitting result of the median width would be altered to

\begin{equation}
\begin{aligned}
\frac{\rm{D_{fil}}}{\rm{Mpc}/h}=0.861+0.279\times log_{10}(\eta_{fil,10})+\\0.177\times [log_{10}(\eta_{fil,10})]^2. 
\label{eqn:dfil_eta_long}
\end{aligned}
\end{equation}

In this case, $56.0\%$ of the grid cells in filament segments longer than 2 $\rm{Mpc}/h$ lie within the range of values predicted by eqn. \ref{eqn:dfil_eta_long} plus or minus 0.5 Mpc/h. 

Alternatively, increasing the neighbor finding radius or removing the restriction that only includes grids and halos within the same filament segments can lead to a moderate reduction in the scatter observed in the $\eta_{fil}$-$D_{fil}$ relation. The results of these adjustments can be found in the bottom left and right panels of Figure \ref{fig:filb_halo_width}. For a neighbor finding radius of $R_f=3.0\, \rm{Mpc}/h$, fitting the median width with a second-order polynomial yields a set of coefficients (2.144, -0.979, 0.480). When all neighboring grids and halos within filaments and nodes are considered, the corresponding coefficients are (1.277, -0.534, 0.471). In both cases, around $52\%$ of the grid cells in filaments fall within the range defined by the fitting curve plus or minus 0.5 Mpc/h. The coefficients of the fitting results are listed in Table 1. 

\begin{figure}[htb]
\begin{center}
\hspace{-0.0cm}
\includegraphics[width=0.45\textwidth, trim=60 360 20 80, clip]{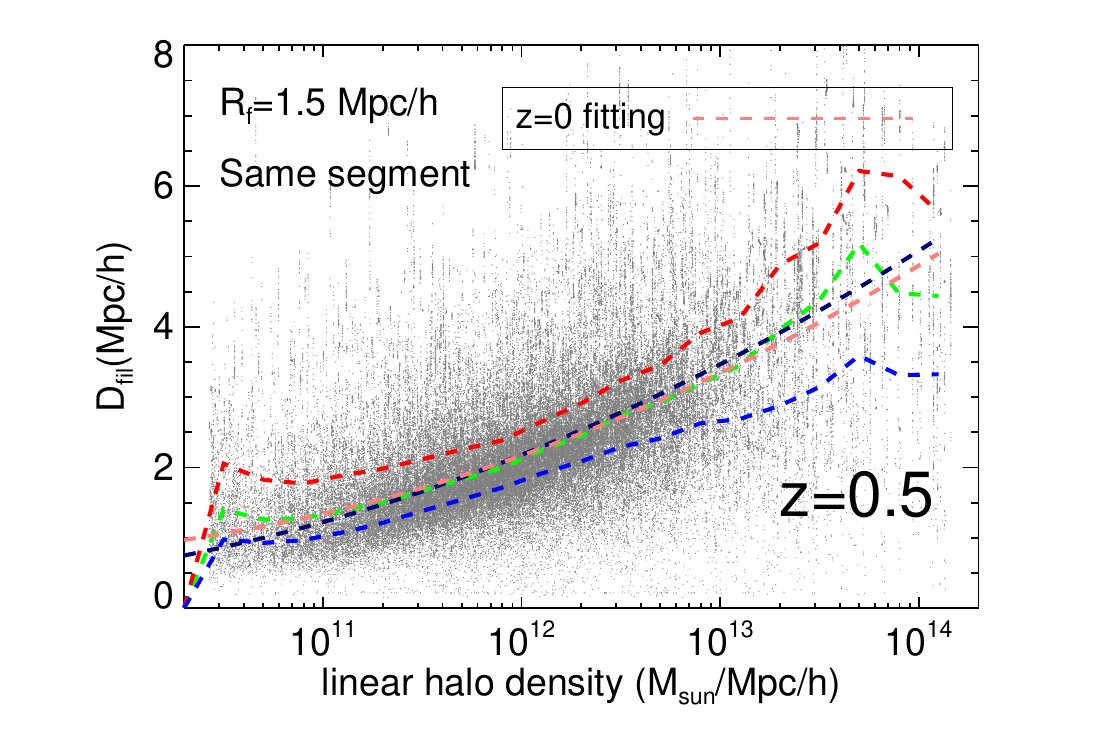}
\includegraphics[width=0.45\textwidth, trim=60 360 20 80, clip]{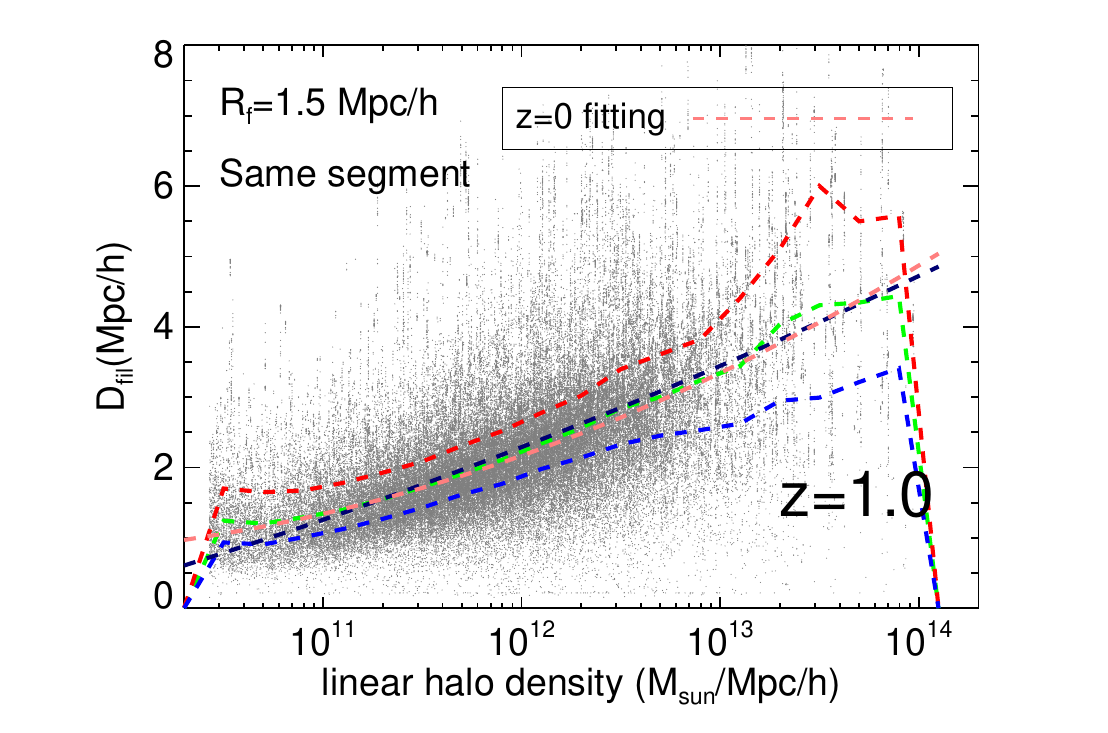}
\caption{The linear halo density and local thickness are estimated at 100,000 grids in filaments. The upper and bottom panels show the results at redshift $z=0.5$ and $1.0$, respectively. The meaning of lines is the same as Figure \ref{fig:filb_halo_width}. Pink line indicates the fitting result at redshift z=0.}
\label{fig:filb_halo_width_hz}
\end{center}
\end{figure}

\begin{deluxetable}{ccccc}[htbp]
\tablenum{1}
\tablecaption{The coefficients obtained from fitting the relation between linear halo density $\eta_{fil}$ and local width ${\rm{D_{fil}}}$ with a second-order polynomial as follows: ${\rm{D_{fil}}}/(\rm{Mpc}/h)=a+b\times log_{10}(\eta_{fil,10})+c\times [log_{10}(\eta_{fil,10})]^2$, where $\eta_{fil,10}=\eta_{fil}/(10^{10}\, \rm{M_\odot/Mpc}/h)$. The `default' case represents the following selection: (1)$\lambda_{th}=0.2$, $R_f=1.5\,\rm{Mpc}/h$; (2) only the neighboring grids and halos in the same segment are considered when estimating $\eta_{fil}$ and $D_{fil}$; (3) all the filament segments, including both longer and shorter than $L_{seg}=2.0 \rm{Mpc}/h$, are considered for the fitting procedure. Other cases are named according to the differences relative to the `default' case.}
\label{tab:den_beta}
\tablewidth{0pt}
\tablehead{
\colhead{redshift} & \colhead{case} & \colhead{a} & \colhead{b} &\colhead{c}}
\startdata
0& default & 0.865 & 0.301 & 0.175 \\
0& $L_{seg}>2.0\, \rm{Mpc}/h$ &0.861 & 0.279 & 0.177 \\
0 & $\rm{R_f}=3.0\, \rm{Mpc}/h$ & 2.144 & -0.979 & 0.480\\
0& All neighbour &1.277 & -0.534 & 0.471 \\
0.5& default &0.596 & 0.463 & 0.164 \\
1.0& default & 0.348 & 0.848 & 0.061 \\
0& $\lambda_{th}=0.4$ & 0.811 & 0.312 & 0.164 \\
\enddata
\end{deluxetable}

\subsection{Results at redshift $z>0$}
We further explore the correlation between the linear halo mass density and filament width at redshift above 0. Figure \ref{fig:filb_halo_width_hz} illustrates the $\eta_{fil}$-$D_{fil}$ relation at redshift $z=0.5$ and $z=1.0$, utilizing a neighbor search radius of $R_f=1.5\, \rm{Mpc}/h$. The fitting result of the `default' case at $z=0$ is also shown in Figure \ref{fig:filb_halo_width_hz}. As the redshift increases, the typical width of filaments decreases due to a reduction in the number of grids within thick filaments, aligning with the anticipated evolution of the cosmic web.  Fitting the median value of the width, a second-order polynomial yields a set of coefficients (0.596, 0.463, 0.164) and (0.348, 0.848, 0.061) at $z=0.5$ and $z=1.0$, respectively. We observe a slight change in the median width at a constant linear halo mass density from z=0 to z=1. Meanwhile, the scatter in the $\eta_{fil}$-$D_{fil}$ relation declines with increasing redshift. $56\%$ and $58\%$ of the grid cells within filaments fall within the range defined by the fitting curves plus or minus 0.5 Mpc/h, at $z=0.5$ and $z=1.0$, respectively. One potential explanation for the decrease in scatter is that tenuous filaments typically exhibit a tighter scaling relation of $\eta_{fil}$-$D_{fil}$, while the number density of thick filaments decreases gradually as redshift increases (\citealt{2021ApJ...920....2Z}). 

\section{Discussions}
\label{sec:discussions}
\subsection{Impact of the cosmic web classification method and further improvement}
\begin{figure*}[htb]
\begin{center}
\hspace{-0.0cm}
\includegraphics[width=0.45\textwidth, trim=60 360 20 80, clip]{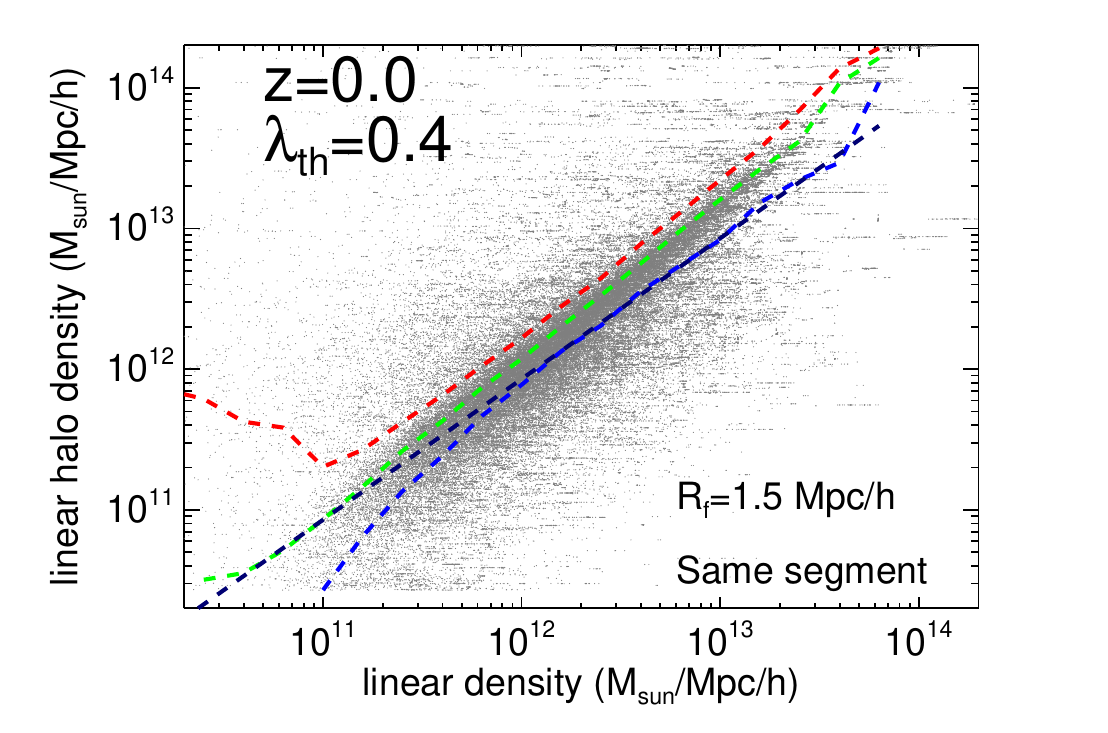}
\includegraphics[width=0.45\textwidth, trim=60 360 20 80, clip]{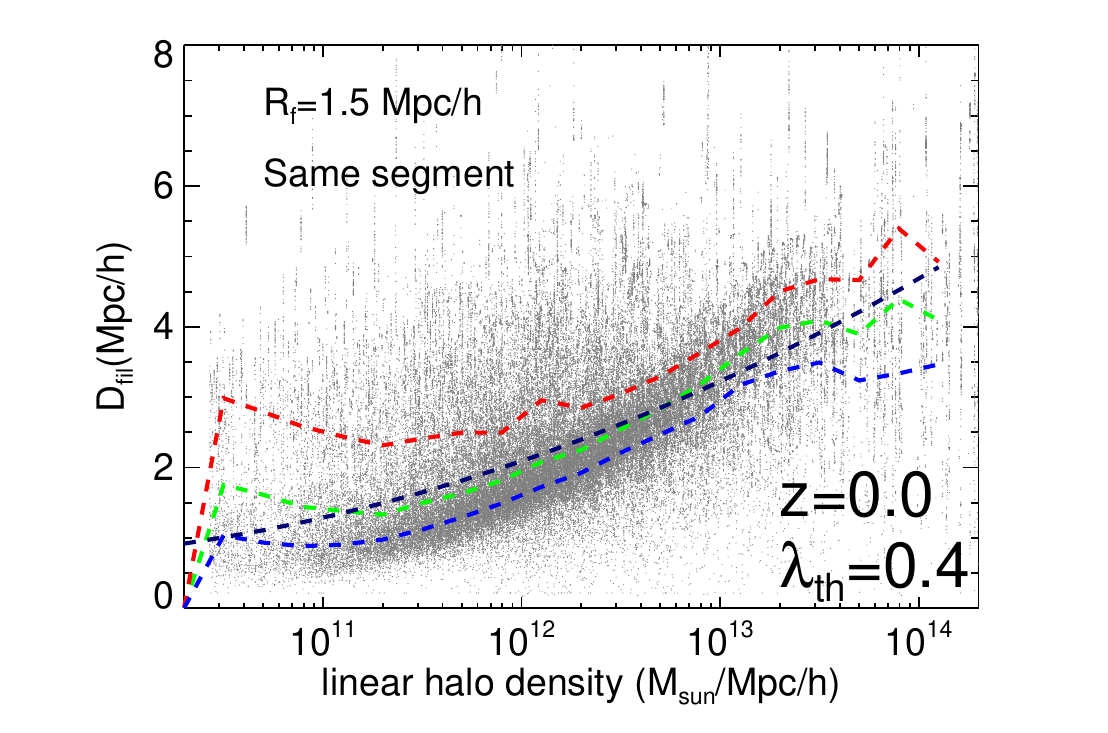}
\caption{The relations between linear halo mass density and linear density (Left), and local width (Right) with $\lambda_{th}=0.4$. The meaning of lines is the same as Figure \ref{fig:filb_halo_width}.}
\label{fig:filb4_relation}
\end{center}
\end{figure*}

At present, there is no standard definition of the cosmic filament; thus, the properties measured for filaments depend on the method and associated parameters used to classify the cosmic web (e.g., \citealt{2018MNRAS.473.1195L}). While the T-web method we utilize is among the most commonly employed algorithms for identifying the cosmic web from simulation, the reliability of the results presented in this study may still be influenced by the procedures and parameters employed for processing filaments. Furthermore, variations in the definitions of halos and halo mass could also affect our findings. For instance, it is valuable to examine how the chosen value of $\lambda_{th}$ affects the findings presented in the preceding section. Figure \ref{fig:filb4_relation} shows the relations between the linear halo mass density and the linear mass density, and local width with $\lambda_{th}=0.4$. Overall, the adjustment on $\lambda_{th}$ has a minor impact on these two relations. The corresponding coefficients for fitting the median width as a function of linear halo mass density with a second-order polynomial are (0.811, 0.312, 0.164). Moreover, there is a slight change in the scatter observed in the relationships $\zeta_{fil}$-$\eta_{fil}$ and $\eta_{fil}$-$D_{fil}$.

While the examination with various $\lambda_{th}$ is somewhat reassuring, we would like to further assert that the scaling relations outlined in this work are generally plausible. With a distinct filament detection method from the T-Web, \citealt{2019MNRAS.486..981G} demonstrated that the total mass of the filaments scales proportionally with the total mass of resident halos, as anticipated by the current theoretical framework of structure formation. It is rational to anticipate a correlation between the linear halo density and the linear density of filament segments. Furthermore, it is anticipated that larger halos are connected by prominent filaments. Consequently, the local width of filaments should generally scale with the mass of embedded halos per unit length. Nevertheless, further research is required to find methods that can minimize the scatter in the relations $\zeta_{fil}$-$\eta_{fil}$ and $\eta_{fil}$-$D_{fil}$. Adjustments in parameters, such as the radius utilized in filament compression and segmentation, might be helpful. Simultaneously, utilizing the total mass of sub-haloes instead of halos may offer a more direct approach to estimating filament width in galaxy samples, as the mass of sub-haloes is more closely linked to the stellar mass.   

Another concern would be how to extrapolate our findings to filaments identified by other methods, especially for samples constructed from halos or galaxies. In fact, \citealt{2018MNRAS.473.1195L} demonstrates that many properties of the cosmic web that are identified through the $\rm{T-Web}$ method are generally consistent with the Disperse (\citealt{2011MNRAS.414..350S}) method, which is usually applied to samples constructed from halos or galaxies. For example, the density probability distribution function in different structures exhibits similarities between the two methods. Furthermore, the volume and mass fractions in knots and filaments categorized by the T-Web broadly align with the mass fraction in filaments identified by Disperse, which lacks the knot structure. This alignment could be attributed in part to both methods indirectly using information from the density field gradient. Additionally, the parameters in both methods have been tuned to capture the visual impression of the cosmic web. Therefore, a straightforward and naive attempt could be employing Disperse to construct the filament network from halo/galaxies data in simulations and subsequently measure the local halo mass density, which can then be utilized to infer the local width of filament. Furthermore, grid cells, mass, and profiles within the filament boundary can be used to compare with the results obtained from the cosmic web classified by the $\rm{T-Web}$ method. If convergence is achieved, the relations can be further used to observed galaxy samples, whose (sub)halo mass can be estimated through the stellar mass - halo mass relation (SHMR). 

To establish a closer match to observed samples of galaxies, we are going to extend the investigation in this study to examine the correlation between the local width of filaments and mass of haloes, sub-haloes, and stellar components using the Illustris-TNG simulation(\citealt{2018MNRAS.475..648P}). This simulation integrates AGN feedback (not present in our current simulation) and has been calibrated to replicate the observed stellar mass function of galaxies.

\subsection{Potential application}
The correlation between the local width and linear halo mass density of the filaments can help to reveal the role of cosmic filament in galaxy evolution and uncover the missing baryons at redshift $z<2$. Filament thickness can exhibit significant variations along the spines (e.g., \citealt{2014MNRAS.441.2923C,2019MNRAS.486..981G}). Filament segments with different thicknesses are likely to have different levels of impact on embedded galaxies, with thick filaments tending to exert different influences on embedded galaxies, with thicker filaments typically possessing greater potential to inhibit gas accretion onto less massive halos and induce more pronounced environmental quenching (\citealt{2019OJAp....2E...7A,2022ApJ...924..132Z,2023arXiv231101443H}). Measurement of the local width allows for examination of how filament segments with different thicknesses impact the evolution of galaxies.

Next, we would like to pay more attention to the `missing' baryons at $z<2$, which are believed to be hosted by filaments and sheets and are the targets of many current and future observational plans covering multiple tools such as X-rays, SZ and Fast Radio Bursts. On the basis of the $D_{fil}$-$\eta_{fil}$ relation, one can deduce the local width for the filaments constructed from observed galaxy distribution. Additionally, the density and temperature profiles have been revealed in the literature (e.g., \citealt{2019MNRAS.486..981G,2021A&A...649A.117G,2021A&A...646A.156T,2021ApJ...906...95Z,2023arXiv230603966L}). For instance, in our previous study \cite{2021ApJ...906...95Z}, we found that the density profiles in the cross-section of filaments exhibit self-similarity and can be described by an isothermal single beta model. By combining the information on width and profiles, we can estimate the mass of dark and baryonic matter residing in the filament segments and predict the associated X-ray and SZ signals, which can then be compared with observations. 

\section{Summary} 
\label{sec:summary}
The width/thickness is an important property of cosmic filaments, which is crucial for the exploration of the role of filaments in galaxy formation and for the detection of missing baryons. In this work, we have investigated the correlation between the local width of cosmic filaments and the mass of dark matter halos within filaments per unit length based on a cosmological hydrodynamic simulation. We find that the local width of filament segments grows with the linear halo mass density increases, which can be approximately described as a second-order polynomial despite some notable scatter. 

The results presented in this work call for further refinement and validation through the utilization of additional simulation samples and the enhancement of the methodologies employed for defining, compressing, and segmenting filaments. Moreover, the methods used to measure the properties of filament could be refined to reduce scatter in the relation between the local width and the linear halo mass density relation. Furthermore, to make our results more conducive to scientific problems related to cosmic filaments, it is essential to implement the findings from this study on filaments constructed from observed galaxy samples. In order to reassess the roles of cosmic filament in shaping the galaxy's properties and estimate the mass density of baryons in filaments based on X-ray and SZ observations, we plan to conduct a relevant study in the near future. 

\acknowledgments
We thank the referee Miguel Aragón for his very useful comments and suggestions that improved the manuscript
This work is supported by the National Natural Science Foundation of China (NFSC) through grant 11733010. W.S.Z. is supported by NSFC grant 12173102. Z.F.P. is supported by NSFC grants 12273006 and 2021A1515012373 from the Natural Science Foundation of Guangdong Province. Y.Z. acknowledges the support from the National Natural Science Foundation of China (NFSC) through grant 12203107 and the Guangdong Basic and Applied Basic Research Foundation with No.2019A1515111098. The cosmological hydrodynamic simulation was performed on the Tianhe-II supercomputer. The post-simulation analysis carried out in this work was completed on the HPC facility of the School of Physics and Astronomy, Sun, Yat-Sen University.

%




\vspace{15mm}




\bibliography{main}{}

\begin{thebibliography}{}
\expandafter\ifx\csname natexlab\endcsname\relax\def\natexlab#1{#1}\fi
\providecommand{\url}[1]{\href{#1}{#1}}
\providecommand{\dodoi}[1]{doi:~\href{http://doi.org/#1}{\nolinkurl{#1}}}
\providecommand{\doeprint}[1]{\href{http://ascl.net/#1}{\nolinkurl{http://ascl.net/#1}}}
\providecommand{\doarXiv}[1]{\href{https://arxiv.org/abs/#1}{\nolinkurl{https://arxiv.org/abs/#1}}}

\bibitem[{{Akamatsu} {et~al.}(2017){Akamatsu}, {Fujita}, {Akahori}, {Ishisaki},
  {Hayashida}, {Hoshino}, {Mernier}, {Yoshikawa}, {Sato}, \&
  {Kaastra}}]{2017A&A...606A...1A}
{Akamatsu}, H., {Fujita}, Y., {Akahori}, T., {et~al.} 2017, \aap, 606, A1,
  \dodoi{10.1051/0004-6361/201730497}

\bibitem[{{Alpaslan} {et~al.}(2014){Alpaslan}, {Robotham}, {Driver}, {Norberg},
  {Baldry}, {Bauer}, {Bland-Hawthorn}, {Brown}, {Cluver}, {Colless}, {Foster},
  {Hopkins}, {Van Kampen}, {Kelvin}, {Lara-Lopez}, {Liske}, {Lopez-Sanchez},
  {Loveday}, {McNaught-Roberts}, {Merson}, \& {Pimbblet}}]{2014MNRAS.438..177A}
{Alpaslan}, M., {Robotham}, A. S.~G., {Driver}, S., {et~al.} 2014, \mnras, 438,
  177, \dodoi{10.1093/mnras/stt2136}

\bibitem[{{Arag{\'o}n-Calvo} {et~al.}(2007){Arag{\'o}n-Calvo}, {Jones}, {van de
  Weygaert}, \& {van der Hulst}}]{2007A&A...474..315A}
{Arag{\'o}n-Calvo}, M.~A., {Jones}, B.~J.~T., {van de Weygaert}, R., \& {van
  der Hulst}, J.~M. 2007, \aap, 474, 315, \dodoi{10.1051/0004-6361:20077880}

\bibitem[{{Aragon Calvo} {et~al.}(2019){Aragon Calvo}, {Neyrinck}, \&
  {Silk}}]{2019OJAp....2E...7A}
{Aragon Calvo}, M.~A., {Neyrinck}, M.~C., \& {Silk}, J. 2019, The Open Journal
  of Astrophysics, 2, 7, \dodoi{10.21105/astro.1697.07881}

\bibitem[{{Arag{\'o}n-Calvo} {et~al.}(2010{\natexlab{a}}){Arag{\'o}n-Calvo},
  {Platen}, {van de Weygaert}, \& {Szalay}}]{2010ApJ...723..364A}
{Arag{\'o}n-Calvo}, M.~A., {Platen}, E., {van de Weygaert}, R., \& {Szalay},
  A.~S. 2010{\natexlab{a}}, \apj, 723, 364, \dodoi{10.1088/0004-637X/723/1/364}

\bibitem[{{Arag{\'o}n-Calvo} {et~al.}(2010{\natexlab{b}}){Arag{\'o}n-Calvo},
  {van de Weygaert}, \& {Jones}}]{2010MNRAS.408.2163A}
{Arag{\'o}n-Calvo}, M.~A., {van de Weygaert}, R., \& {Jones}, B. J.~T.
  2010{\natexlab{b}}, \mnras, 408, 2163,
  \dodoi{10.1111/j.1365-2966.2010.17263.x}

\bibitem[{{Bonamente} {et~al.}(2016){Bonamente}, {Nevalainen}, {Tilton},
  {Liivam{\"a}gi}, {Tempel}, {Hein{\"a}m{\"a}ki}, \&
  {Fang}}]{2016MNRAS.457.4236B}
{Bonamente}, M., {Nevalainen}, J., {Tilton}, E., {et~al.} 2016, \mnras, 457,
  4236, \dodoi{10.1093/mnras/stw285}

\bibitem[{{Bond} {et~al.}(1996){Bond}, {Kofman}, \&
  {Pogosyan}}]{1996Natur.380..603B}
{Bond}, J.~R., {Kofman}, L., \& {Pogosyan}, D. 1996, \nat, 380, 603,
  \dodoi{10.1038/380603a0}

\bibitem[{{Bond} {et~al.}(2010){Bond}, {Strauss}, \&
  {Cen}}]{2010MNRAS.409..156B}
{Bond}, N.~A., {Strauss}, M.~A., \& {Cen}, R. 2010, \mnras, 409, 156,
  \dodoi{10.1111/j.1365-2966.2010.17307.x}

\bibitem[{{Bonjean} {et~al.}(2020){Bonjean}, {Aghanim}, {Douspis}, {Malavasi},
  \& {Tanimura}}]{2020A&A...638A..75B}
{Bonjean}, V., {Aghanim}, N., {Douspis}, M., {Malavasi}, N., \& {Tanimura}, H.
  2020, \aap, 638, A75, \dodoi{10.1051/0004-6361/201937313}

\bibitem[{{Bonjean} {et~al.}(2018){Bonjean}, {Aghanim}, {Salom{\'e}},
  {Douspis}, \& {Beelen}}]{2018A&A...609A..49B}
{Bonjean}, V., {Aghanim}, N., {Salom{\'e}}, P., {Douspis}, M., \& {Beelen}, A.
  2018, \aap, 609, A49, \dodoi{10.1051/0004-6361/201731699}

\bibitem[{{Bregman}(2007)}]{2007ARA&A..45..221B}
{Bregman}, J.~N. 2007, \araa, 45, 221,
  \dodoi{10.1146/annurev.astro.45.051806.110619}

\bibitem[{{Bregman} {et~al.}(2009){Bregman}, {Otte}, {Irwin}, {Putman},
  {Lloyd-Davies}, \& {Br{\"u}ns}}]{2009ApJ...699.1765B}
{Bregman}, J.~N., {Otte}, B., {Irwin}, J.~A., {et~al.} 2009, \apj, 699, 1765,
  \dodoi{10.1088/0004-637X/699/2/1765}

\bibitem[{{Cautun} {et~al.}(2014){Cautun}, {van de Weygaert}, {Jones}, \&
  {Frenk}}]{2014MNRAS.441.2923C}
{Cautun}, M., {van de Weygaert}, R., {Jones}, B. J.~T., \& {Frenk}, C.~S. 2014,
  \mnras, 441, 2923, \dodoi{10.1093/mnras/stu768}

\bibitem[{{Cen} \& {Ostriker}(1999)}]{1999ApJ...514....1C}
{Cen}, R., \& {Ostriker}, J.~P. 1999, \apj, 514, 1, \dodoi{10.1086/306949}

\bibitem[{{Chen} {et~al.}(2017){Chen}, {Ho}, {Mandelbaum}, {Bahcall},
  {Brownstein}, {Freeman}, {Genovese}, {Schneider}, \&
  {Wasserman}}]{2017MNRAS.466.1880C}
{Chen}, Y.-C., {Ho}, S., {Mandelbaum}, R., {et~al.} 2017, \mnras, 466, 1880,
  \dodoi{10.1093/mnras/stw3127}

\bibitem[{{Codis} {et~al.}(2018){Codis}, {Pogosyan}, \&
  {Pichon}}]{2018MNRAS.479..973C}
{Codis}, S., {Pogosyan}, D., \& {Pichon}, C. 2018, \mnras, 479, 973,
  \dodoi{10.1093/mnras/sty1643}

\bibitem[{{Colberg} {et~al.}(2005){Colberg}, {Krughoff}, \&
  {Connolly}}]{2005MNRAS.359..272C}
{Colberg}, J.~M., {Krughoff}, K.~S., \& {Connolly}, A.~J. 2005, \mnras, 359,
  272, \dodoi{10.1111/j.1365-2966.2005.08897.x}

\bibitem[{{Colless} {et~al.}(2003){Colless}, {Peterson}, {Jackson}, {Peacock},
  {Cole}, {Norberg}, {Baldry}, {Baugh}, {Bland-Hawthorn}, {Bridges}, {Cannon},
  {Collins}, {Couch}, {Cross}, {Dalton}, {De Propris}, {Driver}, {Efstathiou},
  {Ellis}, {Frenk}, {Glazebrook}, {Lahav}, {Lewis}, {Lumsden}, {Maddox},
  {Madgwick}, {Sutherland}, \& {Taylor}}]{2003astro.ph..6581C}
{Colless}, M., {Peterson}, B.~A., {Jackson}, C., {et~al.} 2003, arXiv e-prints,
  astro, \dodoi{10.48550/arXiv.astro-ph/0306581}

\bibitem[{{Cui} {et~al.}(2018){Cui}, {Knebe}, {Yepes}, {Yang}, {Borgani},
  {Kang}, {Power}, \& {Staveley-Smith}}]{2018MNRAS.473...68C}
{Cui}, W., {Knebe}, A., {Yepes}, G., {et~al.} 2018, \mnras, 473, 68,
  \dodoi{10.1093/mnras/stx2323}

\bibitem[{{Danforth} {et~al.}(2016){Danforth}, {Keeney}, {Tilton}, {Shull},
  {Stocke}, {Stevans}, {Pieri}, {Savage}, {France}, {Syphers}, {Smith},
  {Green}, {Froning}, {Penton}, \& {Osterman}}]{2016ApJ...817..111D}
{Danforth}, C.~W., {Keeney}, B.~A., {Tilton}, E.~M., {et~al.} 2016, \apj, 817,
  111, \dodoi{10.3847/0004-637X/817/2/111}

\bibitem[{{Dav{\'e}} {et~al.}(2001){Dav{\'e}}, {Cen}, {Ostriker}, {Bryan},
  {Hernquist}, {Katz}, {Weinberg}, {Norman}, \& {O'Shea}}]{2001ApJ...552..473D}
{Dav{\'e}}, R., {Cen}, R., {Ostriker}, J.~P., {et~al.} 2001, \apj, 552, 473,
  \dodoi{10.1086/320548}

\bibitem[{{de Graaff} {et~al.}(2019){de Graaff}, {Cai}, {Heymans}, \&
  {Peacock}}]{2019A&A...624A..48D}
{de Graaff}, A., {Cai}, Y.-C., {Heymans}, C., \& {Peacock}, J.~A. 2019, \aap,
  624, A48, \dodoi{10.1051/0004-6361/201935159}

\bibitem[{{de Lapparent} {et~al.}(1986){de Lapparent}, {Geller}, \&
  {Huchra}}]{1986ApJ...302L...1D}
{de Lapparent}, V., {Geller}, M.~J., \& {Huchra}, J.~P. 1986, \apjl, 302, L1,
  \dodoi{10.1086/184625}

\bibitem[{{Dolag} {et~al.}(2006){Dolag}, {Meneghetti}, {Moscardini}, {Rasia},
  \& {Bonaldi}}]{2006MNRAS.370..656D}
{Dolag}, K., {Meneghetti}, M., {Moscardini}, L., {Rasia}, E., \& {Bonaldi}, A.
  2006, \mnras, 370, 656, \dodoi{10.1111/j.1365-2966.2006.10511.x}

\bibitem[{{Eckert} {et~al.}(2015){Eckert}, {Jauzac}, {Shan}, {Kneib}, {Erben},
  {Israel}, {Jullo}, {Klein}, {Massey}, {Richard}, \&
  {Tchernin}}]{2015Natur.528..105E}
{Eckert}, D., {Jauzac}, M., {Shan}, H., {et~al.} 2015, \nat, 528, 105,
  \dodoi{10.1038/nature16058}

\bibitem[{{Fang} {et~al.}(2002){Fang}, {Marshall}, {Lee}, {Davis}, \&
  {Canizares}}]{2002ApJ...572L.127F}
{Fang}, T., {Marshall}, H.~L., {Lee}, J.~C., {Davis}, D.~S., \& {Canizares},
  C.~R. 2002, \apjl, 572, L127, \dodoi{10.1086/341665}

\bibitem[{{Forero-Romero} {et~al.}(2009){Forero-Romero}, {Hoffman},
  {Gottl{\"o}ber}, {Klypin}, \& {Yepes}}]{2009MNRAS.396.1815F}
{Forero-Romero}, J.~E., {Hoffman}, Y., {Gottl{\"o}ber}, S., {Klypin}, A., \&
  {Yepes}, G. 2009, \mnras, 396, 1815, \dodoi{10.1111/j.1365-2966.2009.14885.x}

\bibitem[{{Fukugita} {et~al.}(1998){Fukugita}, {Hogan}, \&
  {Peebles}}]{1998ApJ...503..518F}
{Fukugita}, M., {Hogan}, C.~J., \& {Peebles}, P.~J.~E. 1998, \apj, 503, 518,
  \dodoi{10.1086/306025}

\bibitem[{{Gal{\'a}rraga-Espinosa} {et~al.}(2021){Gal{\'a}rraga-Espinosa},
  {Aghanim}, {Langer}, \& {Tanimura}}]{2021A&A...649A.117G}
{Gal{\'a}rraga-Espinosa}, D., {Aghanim}, N., {Langer}, M., \& {Tanimura}, H.
  2021, \aap, 649, A117, \dodoi{10.1051/0004-6361/202039781}

\bibitem[{{Gheller} \& {Vazza}(2019)}]{2019MNRAS.486..981G}
{Gheller}, C., \& {Vazza}, F. 2019, \mnras, 486, 981,
  \dodoi{10.1093/mnras/stz843}

\bibitem[{{Gheller} {et~al.}(2015){Gheller}, {Vazza}, {Favre}, \&
  {Br{\"u}ggen}}]{2015MNRAS.453.1164G}
{Gheller}, C., {Vazza}, F., {Favre}, J., \& {Br{\"u}ggen}, M. 2015, \mnras,
  453, 1164, \dodoi{10.1093/mnras/stv1646}

\bibitem[{{Gonz{\'a}lez} \& {Padilla}(2010)}]{2010MNRAS.407.1449G}
{Gonz{\'a}lez}, R.~E., \& {Padilla}, N.~D. 2010, \mnras, 407, 1449,
  \dodoi{10.1111/j.1365-2966.2010.17015.x}

\bibitem[{{Haardt} \& {Madau}(1996)}]{1996ApJ...461...20H}
{Haardt}, F., \& {Madau}, P. 1996, \apj, 461, 20, \dodoi{10.1086/177035}

\bibitem[{{Hahn} {et~al.}(2007){Hahn}, {Porciani}, {Carollo}, \&
  {Dekel}}]{2007MNRAS.375..489H}
{Hahn}, O., {Porciani}, C., {Carollo}, C.~M., \& {Dekel}, A. 2007, \mnras, 375,
  489, \dodoi{10.1111/j.1365-2966.2006.11318.x}

\bibitem[{{Haider} {et~al.}(2016){Haider}, {Steinhauser}, {Vogelsberger},
  {Genel}, {Springel}, {Torrey}, \& {Hernquist}}]{2016MNRAS.457.3024H}
{Haider}, M., {Steinhauser}, D., {Vogelsberger}, M., {et~al.} 2016, \mnras,
  457, 3024, \dodoi{10.1093/mnras/stw077}

\bibitem[{{Hasan} {et~al.}(2023){Hasan}, {Burchett}, {Hellinger}, {Elek},
  {Nagai}, {Faber}, {Primack}, {Koo}, {Mandelker}, \&
  {Woo}}]{2023arXiv231101443H}
{Hasan}, F., {Burchett}, J.~N., {Hellinger}, D., {et~al.} 2023, arXiv e-prints,
  arXiv:2311.01443, \dodoi{10.48550/arXiv.2311.01443}

\bibitem[{{Ho} {et~al.}(2018){Ho}, {Gronke}, {Falck}, \&
  {Mota}}]{2018A&A...619A.122H}
{Ho}, A., {Gronke}, M., {Falck}, B., \& {Mota}, D.~F. 2018, \aap, 619, A122,
  \dodoi{10.1051/0004-6361/201833899}

\bibitem[{{Hoffman} {et~al.}(2012){Hoffman}, {Metuki}, {Yepes},
  {Gottl{\"o}ber}, {Forero-Romero}, {Libeskind}, \&
  {Knebe}}]{2012MNRAS.425.2049H}
{Hoffman}, Y., {Metuki}, O., {Yepes}, G., {et~al.} 2012, \mnras, 425, 2049,
  \dodoi{10.1111/j.1365-2966.2012.21553.x}

\bibitem[{{Kraljic} {et~al.}(2018){Kraljic}, {Arnouts}, {Pichon}, {Laigle}, {de
  la Torre}, {Vibert}, {Cadiou}, {Dubois}, {Treyer}, {Schimd}, {Codis}, {de
  Lapparent}, {Devriendt}, {Hwang}, {Le Borgne}, {Malavasi}, {Milliard},
  {Musso}, {Pogosyan}, {Alpaslan}, {Bland-Hawthorn}, \&
  {Wright}}]{2018MNRAS.474..547K}
{Kraljic}, K., {Arnouts}, S., {Pichon}, C., {et~al.} 2018, \mnras, 474, 547,
  \dodoi{10.1093/mnras/stx2638}

\bibitem[{{Kuutma} {et~al.}(2017){Kuutma}, {Tamm}, \&
  {Tempel}}]{2017A&A...600L...6K}
{Kuutma}, T., {Tamm}, A., \& {Tempel}, E. 2017, \aap, 600, L6,
  \dodoi{10.1051/0004-6361/201730526}

\bibitem[{{Lavaux} \& {Wandelt}(2010)}]{2010MNRAS.403.1392L}
{Lavaux}, G., \& {Wandelt}, B.~D. 2010, \mnras, 403, 1392,
  \dodoi{10.1111/j.1365-2966.2010.16197.x}

\bibitem[{{Libeskind} {et~al.}(2018){Libeskind}, {van de Weygaert}, {Cautun},
  {Falck}, {Tempel}, {Abel}, {Alpaslan}, {Arag{\'o}n-Calvo}, {Forero-Romero},
  {Gonzalez}, {Gottl{\"o}ber}, {Hahn}, {Hellwing}, {Hoffman}, {Jones},
  {Kitaura}, {Knebe}, {Manti}, {Neyrinck}, {Nuza}, {Padilla}, {Platen},
  {Ramachandra}, {Robotham}, {Saar}, {Shandarin}, {Steinmetz}, {Stoica},
  {Sousbie}, \& {Yepes}}]{2018MNRAS.473.1195L}
{Libeskind}, N.~I., {van de Weygaert}, R., {Cautun}, M., {et~al.} 2018, \mnras,
  473, 1195, \dodoi{10.1093/mnras/stx1976}

\bibitem[{{Lu} {et~al.}(2023){Lu}, {Mandelker}, {Oh}, {Dekel}, {van den Bosch},
  {Springel}, {Nagai}, \& {van de Voort}}]{2023arXiv230603966L}
{Lu}, Y.~S., {Mandelker}, N., {Oh}, S.~P., {et~al.} 2023, arXiv e-prints,
  arXiv:2306.03966, \dodoi{10.48550/arXiv.2306.03966}

\bibitem[{{Martizzi} {et~al.}(2019){Martizzi}, {Vogelsberger}, {Artale},
  {Haider}, {Torrey}, {Marinacci}, {Nelson}, {Pillepich}, {Weinberger},
  {Hernquist}, {Naiman}, \& {Springel}}]{2019MNRAS.486.3766M}
{Martizzi}, D., {Vogelsberger}, M., {Artale}, M.~C., {et~al.} 2019, \mnras,
  486, 3766, \dodoi{10.1093/mnras/stz1106}

\bibitem[{{Musso} {et~al.}(2018){Musso}, {Cadiou}, {Pichon}, {Codis},
  {Kraljic}, \& {Dubois}}]{2018MNRAS.476.4877M}
{Musso}, M., {Cadiou}, C., {Pichon}, C., {et~al.} 2018, \mnras, 476, 4877,
  \dodoi{10.1093/mnras/sty191}

\bibitem[{{Nevalainen} {et~al.}(2019){Nevalainen}, {Tempel}, {Ahoranta},
  {Liivam{\"a}gi}, {Bonamente}, {Tilton}, {Kaastra}, {Fang},
  {Hein{\"a}m{\"a}ki}, {Saar}, \& {Finoguenov}}]{2019A&A...621A..88N}
{Nevalainen}, J., {Tempel}, E., {Ahoranta}, J., {et~al.} 2019, \aap, 621, A88,
  \dodoi{10.1051/0004-6361/201833109}

\bibitem[{{Nicastro} {et~al.}(2005){Nicastro}, {Mathur}, {Elvis}, {Drake},
  {Fiore}, {Fang}, {Fruscione}, {Krongold}, {Marshall}, \&
  {Williams}}]{2005ApJ...629..700N}
{Nicastro}, F., {Mathur}, S., {Elvis}, M., {et~al.} 2005, \apj, 629, 700,
  \dodoi{10.1086/431270}

\bibitem[{{Nicastro} {et~al.}(2018){Nicastro}, {Kaastra}, {Krongold},
  {Borgani}, {Branchini}, {Cen}, {Dadina}, {Danforth}, {Elvis}, {Fiore},
  {Gupta}, {Mathur}, {Mayya}, {Paerels}, {Piro}, {Rosa-Gonzalez}, {Schaye},
  {Shull}, {Torres-Zafra}, {Wijers}, \& {Zappacosta}}]{2018Natur.558..406N}
{Nicastro}, F., {Kaastra}, J., {Krongold}, Y., {et~al.} 2018, \nat, 558, 406,
  \dodoi{10.1038/s41586-018-0204-1}

\bibitem[{{Novikov} {et~al.}(2006){Novikov}, {Colombi}, \&
  {Dor{\'e}}}]{2006MNRAS.366.1201N}
{Novikov}, D., {Colombi}, S., \& {Dor{\'e}}, O. 2006, \mnras, 366, 1201,
  \dodoi{10.1111/j.1365-2966.2005.09925.x}

\bibitem[{{Pillepich} {et~al.}(2018){Pillepich}, {Nelson}, {Hernquist},
  {Springel}, {Pakmor}, {Torrey}, {Weinberger}, {Genel}, {Naiman}, {Marinacci},
  \& {Vogelsberger}}]{2018MNRAS.475..648P}
{Pillepich}, A., {Nelson}, D., {Hernquist}, L., {et~al.} 2018, \mnras, 475,
  648, \dodoi{10.1093/mnras/stx3112}

\bibitem[{{Planck Collaboration} {et~al.}(2014){Planck Collaboration}, {Ade},
  {Aghanim}, {Armitage-Caplan}, {Arnaud}, {Ashdown}, {Atrio-Barand ela},
  {Aumont}, {Baccigalupi}, {Banday}, {Barreiro}, {Bartlett}, {Battaner},
  {Benabed}, {Beno{\^\i}t}, {Benoit-L{\'e}vy}, {Bernard}, {Bersanelli},
  {Bielewicz}, {Bobin}, {Bock}, {Bonaldi}, {Bond}, {Borrill}, {Bouchet},
  {Bridges}, {Bucher}, {Burigana}, {Butler}, {Calabrese}, {Cappellini},
  {Cardoso}, {Catalano}, {Challinor}, {Chamballu}, {Chary}, {Chen}, {Chiang},
  {Chiang}, {Christensen}, {Church}, {Clements}, {Colombi}, {Colombo},
  {Couchot}, {Coulais}, {Crill}, {Curto}, {Cuttaia}, {Danese}, {Davies},
  {Davis}, {de Bernardis}, {de Rosa}, {de Zotti}, {Delabrouille}, {Delouis},
  {D{\'e}sert}, {Dickinson}, {Diego}, {Dolag}, {Dole}, {Donzelli}, {Dor{\'e}},
  {Douspis}, {Dunkley}, {Dupac}, {Efstathiou}, {Elsner}, {En{\ss}lin},
  {Eriksen}, {Finelli}, {Forni}, {Frailis}, {Fraisse}, {Franceschi}, {Gaier},
  {Galeotta}, {Galli}, {Ganga}, {Giard}, {Giardino}, {Giraud-H{\'e}raud},
  {Gjerl{\o}w}, {Gonz{\'a}lez-Nuevo}, {G{\'o}rski}, {Gratton}, {Gregorio},
  {Gruppuso}, {Gudmundsson}, {Haissinski}, {Hamann}, {Hansen}, {Hanson},
  {Harrison}, {Henrot-Versill{\'e}}, {Hern{\'a}ndez-Monteagudo}, {Herranz},
  {Hildebrand t}, {Hivon}, {Hobson}, {Holmes}, {Hornstrup}, {Hou}, {Hovest},
  {Huffenberger}, {Jaffe}, {Jaffe}, {Jewell}, {Jones}, {Juvela},
  {Keih{\"a}nen}, {Keskitalo}, {Kisner}, {Kneissl}, {Knoche}, {Knox}, {Kunz},
  {Kurki-Suonio}, {Lagache}, {L{\"a}hteenm{\"a}ki}, {Lamarre}, {Lasenby},
  {Lattanzi}, {Laureijs}, {Lawrence}, {Leach}, {Leahy}, {Leonardi},
  {Le{\'o}n-Tavares}, {Lesgourgues}, {Lewis}, {Liguori}, {Lilje},
  {Linden-V{\o}rnle}, {L{\'o}pez-Caniego}, {Lubin}, {Mac{\'\i}as-P{\'e}rez},
  {Maffei}, {Maino}, {Mand olesi}, {Maris}, {Marshall}, {Martin},
  {Mart{\'\i}nez-Gonz{\'a}lez}, {Masi}, {Massardi}, {Matarrese}, {Matthai},
  {Mazzotta}, {Meinhold}, {Melchiorri}, {Melin}, {Mendes}, {Menegoni},
  {Mennella}, {Migliaccio}, {Millea}, {Mitra}, {Miville-Desch{\^e}nes},
  {Moneti}, {Montier}, {Morgante}, {Mortlock}, {Moss}, {Munshi}, {Murphy},
  {Naselsky}, {Nati}, {Natoli}, {Netterfield}, {N{\o}rgaard-Nielsen},
  {Noviello}, {Novikov}, {Novikov}, {O'Dwyer}, {Osborne}, {Oxborrow}, {Paci},
  {Pagano}, {Pajot}, {Paladini}, {Paoletti}, {Partridge}, {Pasian},
  {Patanchon}, {Pearson}, {Pearson}, {Peiris}, {Perdereau}, {Perotto},
  {Perrotta}, {Pettorino}, {Piacentini}, {Piat}, {Pierpaoli}, {Pietrobon},
  {Plaszczynski}, {Platania}, {Pointecouteau}, {Polenta}, {Ponthieu}, {Popa},
  {Poutanen}, {Pratt}, {Pr{\'e}zeau}, {Prunet}, {Puget}, {Rachen}, {Reach},
  {Rebolo}, {Reinecke}, {Remazeilles}, {Renault}, {Ricciardi}, {Riller},
  {Ristorcelli}, {Rocha}, {Rosset}, {Roudier}, {Rowan-Robinson},
  {Rubi{\~n}o-Mart{\'\i}n}, {Rusholme}, {Sandri}, {Santos}, {Savelainen},
  {Savini}, {Scott}, {Seiffert}, {Shellard}, {Spencer}, {Starck}, {Stolyarov},
  {Stompor}, {Sudiwala}, {Sunyaev}, {Sureau}, {Sutton}, {Suur-Uski}, {Sygnet},
  {Tauber}, {Tavagnacco}, {Terenzi}, {Toffolatti}, {Tomasi}, {Tristram},
  {Tucci}, {Tuovinen}, {T{\"u}rler}, {Umana}, {Valenziano}, {Valiviita}, {Van
  Tent}, {Vielva}, {Villa}, {Vittorio}, {Wade}, {Wandelt}, {Wehus}, {White},
  {White}, {Wilkinson}, {Yvon}, {Zacchei}, \& {Zonca}}]{2014A&A...571A..16P}
{Planck Collaboration}, {Ade}, P.~A.~R., {Aghanim}, N., {et~al.} 2014, \aap,
  571, A16, \dodoi{10.1051/0004-6361/201321591}

\bibitem[{{Shull} {et~al.}(2012){Shull}, {Smith}, \&
  {Danforth}}]{2012ApJ...759...23S}
{Shull}, J.~M., {Smith}, B.~D., \& {Danforth}, C.~W. 2012, \apj, 759, 23,
  \dodoi{10.1088/0004-637X/759/1/23}

\bibitem[{{Singh} {et~al.}(2020){Singh}, {Mahajan}, \&
  {Bagla}}]{2020MNRAS.497.2265S}
{Singh}, A., {Mahajan}, S., \& {Bagla}, J.~S. 2020, \mnras, 497, 2265,
  \dodoi{10.1093/mnras/staa1913}

\bibitem[{{Sousbie}(2011)}]{2011MNRAS.414..350S}
{Sousbie}, T. 2011, \mnras, 414, 350, \dodoi{10.1111/j.1365-2966.2011.18394.x}

\bibitem[{{Tanimura} {et~al.}(2020{\natexlab{a}}){Tanimura}, {Aghanim},
  {Bonjean}, {Malavasi}, \& {Douspis}}]{2020A&A...637A..41T}
{Tanimura}, H., {Aghanim}, N., {Bonjean}, V., {Malavasi}, N., \& {Douspis}, M.
  2020{\natexlab{a}}, \aap, 637, A41, \dodoi{10.1051/0004-6361/201937158}

\bibitem[{{Tanimura} {et~al.}(2020{\natexlab{b}}){Tanimura}, {Aghanim},
  {Kolodzig}, {Douspis}, \& {Malavasi}}]{2020A&A...643L...2T}
{Tanimura}, H., {Aghanim}, N., {Kolodzig}, A., {Douspis}, M., \& {Malavasi}, N.
  2020{\natexlab{b}}, \aap, 643, L2, \dodoi{10.1051/0004-6361/202038521}

\bibitem[{{Tanimura} {et~al.}(2019){Tanimura}, {Hinshaw}, {McCarthy}, {Van
  Waerbeke}, {Aghanim}, {Ma}, {Mead}, {Hojjati}, \&
  {Tr{\"o}ster}}]{2019MNRAS.483..223T}
{Tanimura}, H., {Hinshaw}, G., {McCarthy}, I.~G., {et~al.} 2019, \mnras, 483,
  223, \dodoi{10.1093/mnras/sty3118}

\bibitem[{{Tempel} {et~al.}(2016){Tempel}, {Stoica}, {Kipper}, \&
  {Saar}}]{2016A&C....16...17T}
{Tempel}, E., {Stoica}, R.~S., {Kipper}, R., \& {Saar}, E. 2016, Astronomy and
  Computing, 16, 17, \dodoi{10.1016/j.ascom.2016.03.004}

\bibitem[{{Tempel} {et~al.}(2014){Tempel}, {Stoica}, {Mart{\'\i}nez},
  {Liivam{\"a}gi}, {Castellan}, \& {Saar}}]{2014MNRAS.438.3465T}
{Tempel}, E., {Stoica}, R.~S., {Mart{\'\i}nez}, V.~J., {et~al.} 2014, \mnras,
  438, 3465, \dodoi{10.1093/mnras/stt2454}

\bibitem[{{Teyssier}(2002)}]{2002A&A...385..337T}
{Teyssier}, R. 2002, \aap, 385, 337, \dodoi{10.1051/0004-6361:20011817}

\bibitem[{{Tuominen} {et~al.}(2020){Tuominen}, {Nevalainen}, {Tempel},
  {Kuutma}, {Wijers}, {Schaye}, {Hein{\"a}m{\"a}ki}, {Bonamente}, \&
  {Ganeshaiah Veena}}]{2020arXiv201209203T}
{Tuominen}, T., {Nevalainen}, J., {Tempel}, E., {et~al.} 2020, arXiv e-prints,
  arXiv:2012.09203.
\newblock \doarXiv{2012.09203}

\bibitem[{{Tuominen} {et~al.}(2021){Tuominen}, {Nevalainen}, {Tempel},
  {Kuutma}, {Wijers}, {Schaye}, {Hein{\"a}m{\"a}ki}, {Bonamente}, \&
  {Ganeshaiah Veena}}]{2021A&A...646A.156T}
---. 2021, \aap, 646, A156, \dodoi{10.1051/0004-6361/202039221}

\bibitem[{{van de Weygaert} \& {Bond}(2008)}]{2008LNP...740..335V}
{van de Weygaert}, R., \& {Bond}, J.~R. 2008, in A Pan-Chromatic View of
  Clusters of Galaxies and the Large-Scale Structure, ed. M.~{Plionis},
  O.~{L{\'o}pez-Cruz}, \& D.~{Hughes}, Vol. 740, 335,
  \dodoi{10.1007/978-1-4020-6941-3_10}

\bibitem[{{van de Weygaert} \& {Platen}(2011)}]{2011IJMPS...1...41V}
{van de Weygaert}, R., \& {Platen}, E. 2011, in International Journal of Modern
  Physics Conference Series, Vol.~1, International Journal of Modern Physics
  Conference Series, 41--66, \dodoi{10.1142/S2010194511000092}

\bibitem[{{Zakharova} {et~al.}(2023){Zakharova}, {Vulcani}, {De Lucia}, {Xie},
  {Hirschmann}, \& {Fontanot}}]{2023MNRAS.525.4079Z}
{Zakharova}, D., {Vulcani}, B., {De Lucia}, G., {et~al.} 2023, \mnras, 525,
  4079, \dodoi{10.1093/mnras/stad2562}

\bibitem[{{Zel'dovich}(1970)}]{1970A&A.....5...84Z}
{Zel'dovich}, Y.~B. 1970, \aap, 5, 84

\bibitem[{{Zhu} \& {Feng}(2017)}]{2017ApJ...838...21Z}
{Zhu}, W., \& {Feng}, L.-L. 2017, \apj, 838, 21,
  \dodoi{10.3847/1538-4357/aa61f9}

\bibitem[{{Zhu} \& {Feng}(2021)}]{2021ApJ...906...95Z}
---. 2021, \apj, 906, 95, \dodoi{10.3847/1538-4357/abcb90}

\bibitem[{{Zhu} {et~al.}(2021){Zhu}, {Zhang}, \& {Feng}}]{2021ApJ...920....2Z}
{Zhu}, W., {Zhang}, F., \& {Feng}, L.-L. 2021, \apj, 920, 2,
  \dodoi{10.3847/1538-4357/ac15f1}

\bibitem[{{Zhu} {et~al.}(2022){Zhu}, {Zhang}, \& {Feng}}]{2022ApJ...924..132Z}
---. 2022, \apj, 924, 132, \dodoi{10.3847/1538-4357/ac37b9}

\end{thebibliography}
\bibliographystyle{aasjournal}



\end{document}